\documentclass[usenatbib,useAMS]{mnras}
\usepackage{graphicx}	
\usepackage{multicol}	
\usepackage[english]{babel}
\usepackage{epstopdf}

\newcommand{\mpchi}{\,h^{-1}{\rm {Mpc}}}
\newcommand{\kms}{\,{\rm {km\, s^{-1}}}}
\newcommand{\msun}{\,h^{-1}{\rm M_{\sun}}}

\title[Modelling Galaxy Clustering]{Modelling galaxy clustering: halo occupation distribution versus subhalo matching}

\author[H. Guo et al.]{\parbox{\textwidth}{
Hong Guo$^{1,2}$\thanks{E-mail: guohong@shao.ac.cn}, Zheng Zheng$^{2}$, Peter
S. Behroozi$^{3}$\thanks{Hubble Fellow}, Idit Zehavi$^{4,5}$,  Chia-Hsun Chuang$^{6}$\thanks{MultiDark Fellow}, Johan
Comparat$^{6,7}$\thanks{Severo Ochoa IFT Fellow}, Ginevra Favole$^{6,8}$, Stefan Gottloeber$^{9}$, Anatoly
Klypin$^{10,11}$, Francisco Prada$^{6,8,12}$, Sergio A. Rodr\'{i}guez-Torres
$^{6,7,8}$\thanks{Campus de Excelencia Internacional UAM/CSIC Scholar}
, David H. Weinberg$^{13,14}$, and
Gustavo Yepes$^{7}$}
\vspace*{6pt} \\
$^{1}$ Key Laboratory for Research in Galaxies and Cosmology, Shanghai Astronomical Observatory, Shanghai 200030, China\\
$^{2}$ Department of Physics and Astronomy, University of Utah, UT 84112, USA\\
$^{3}$ Astronomy and Physics Departments and Theoretical Astrophysics Center, University of California, Berkeley, CA 94720, USA\\
$^{4}$ Department of Astronomy, Case Western Reserve University, OH 44106, USA\\
$^{5}$ Institute for Computational Cosmology, Department of Physics,
University of Durham, South Road, Durham, DH1 3LE, UK\\
$^{6}$ Instituto de F\'{\i}sica Te\'orica, (UAM/CSIC), Universidad Aut\'onoma de Madrid,  Cantoblanco,
E-28049 Madrid, Spain \\
$^{7}$ Departamento de F{\'i}sica Te{\'o}rica,  Universidad Aut{\'o}noma de Madrid, Cantoblanco, 28049, Madrid, Spain\\
$^{8}$ Campus of International Excellence UAM+CSIC, Cantoblanco, E-28049 Madrid, Spain\\
$^{9}$ Leibniz-Institut fur Astrophysik (AIP), An der Sternwarte 16, D-14482 Potsdam, Germany\\
$^{10}$ Astronomy Department, New Mexico State University, MSC 4500, PO Box 30001, Las Cruces, NM, 880003-8001, USA\\
$^{11}$Severo Ochoa Associate Researcher at the Instituto de Fisica Teorica (UAM/CSIC), Madrid, Spain\\
$^{12}$ Instituto de Astrof\'{\i}sica de Andaluc\'{\i}a (CSIC), Glorieta de
la Astronom\'{\i}a, E-18080 Granada, Spain
\\
$^{13}$ Department of Astronomy, Ohio State University, Columbus, OH 43210, USA\\
$^{14}$ Center for Cosmology and Astro-Particle Physics, Ohio State
University, Columbus, OH 43210, USA}

\begin{document}
\label{firstpage} \pagerange{\pageref{firstpage}--\pageref{lastpage}}
\maketitle

\begin{abstract}
We model the luminosity-dependent projected and redshift-space two-point
correlation functions (2PCFs) of the Sloan Digital Sky Survey (SDSS) Data Relese 7 Main
galaxy sample, using the halo occupation distribution (HOD) model and the
subhalo abundance matching (SHAM) model and its extension. All the models are
built on the same high-resolution $N$-body simulations. We find that the HOD
model generally provides the best performance in reproducing the clustering
measurements in both projected and redshift spaces. The SHAM model with the
same halo-galaxy relation for central and satellite galaxies (or distinct
haloes and subhaloes), when including scatters, has a best-fitting
$\chi^2/\rm{dof}$ around $2$--$3$. We therefore extend the SHAM model to the
subhalo clustering and abundance matching (SCAM) by allowing the central and
satellite galaxies to have different galaxy--halo relations. We infer the
corresponding halo/subhalo parameters by jointly fitting the galaxy 2PCFs and
abundances and consider subhaloes selected based on three properties, the mass
$M_{\rm acc}$ at the time of accretion, the maximum circular velocity $V_{\rm
acc}$ at the time of accretion, and the peak maximum circular velocity
$V_{\rm peak}$ over the history of the subhaloes. The three subhalo models
work well for luminous galaxy samples (with luminosity above $L_*$). For
low-luminosity samples, the $V_{\rm acc}$ model stands out in reproducing the
data, with the $V_{\rm peak}$ model slightly worse, while the $M_{\rm acc}$
model fails to fit the data. We discuss the implications of the modelling
results.
\end{abstract}


\section{Introduction}
The connection between the observed galaxy distribution and the
underlying dark matter is a fundamental question in modern cosmology. It can
help us understand the dark matter component of the energy density
distribution from the observed baryon components. The contemporary galaxy
formation models assume that galaxies form and evolve within the dark matter
haloes \citep{White78}. Therefore, we can use the dark matter haloes to build
the connection between the luminous and dark sides of the universe.

There are multiple ways of linking galaxies to the dark matter haloes. The
most straightforward method is to employ the hydrodynamic simulations to take
into account the complicated physics involved in the galaxy formation and
evolution \citep[see the latest such simulations in
e.g.][]{Vogelsberger14a,Schaye15}, as well as the semi-analytic models that
are built on the halo merger trees from $N$-body dark matter simulations
\citep[e.g.][]{Bower06,Croton06,Somerville08,Guo11}. But the poorly
understood galaxy formation physical processes related to baryons make such
methods model dependent and difficult to satisfactorily reproduce the
observations in the current data accuracy. Other statistical methods are then
developed to evade the necessity of including the galaxy formation physics
and to make use of the population of dark matter haloes whose formation is
dominated by gravity and well understood. Such methods aim at empirically
establishing the connection between galaxies and dark matter haloes from
statistical distributions of galaxies like galaxy clustering, and then the
galaxy-halo connection is used to constrain galaxy formation and evolution.
The most popular models are the halo occupation distribution
\citep[HOD;][]{Jing98,Peacock00,Berlind02,Zheng05,Zheng09,Leauthaud12,Guo14,Skibba15,Zu15},
the closely related conditional luminosity function
\citep[CLF;][]{Yang03,Yang04}, and the subhalo abundance matching
\citep[SHAM;][]{Kravtsov04,Conroy06,Vale06,Wang07,Behroozi10,Guo10,Moster10,
Nuza13,Rodriguez-Puebla13,Sawala15,Yamamoto15}.
All of these methods are based on the halo framework, by assuming that all
galaxies reside in the haloes. In this paper we focus on the detailed and
quantitative model comparisons between the HOD and SHAM methods.

The HOD description includes the probability $P(N|M)$ of finding $N$ galaxies
of certain properties in a dark matter halo of virial mass $M$, and the
spatial and velocity distribution of those galaxies inside haloes. Analytical
methods have been developed within the HOD (or CLF) framework to compute
galaxy clustering statistics \citep[e.g.][]{Zheng04,Tinker05,Bosch13}. By
using dark matter haloes identified in high-resolution $N$-body simulations,
the HOD model can be made accurate enough to interpret the observed
high-precision galaxy clustering measurements from large galaxy surveys
\citep[][]{Zheng16}, which overcomes the difficulty of modelling the effects
of halo exclusion, nonlinear growth, and scale-dependent halo bias in the
analytical HOD models \citep[e.g.][]{Zheng04,Tinker05}. Based on galaxy
formation models, galaxies in the HOD model are further categorized into
central and satellite galaxies according to their spatial distribution within
the haloes. In many applications, central galaxies are usually put at halo
centres and assumed to have the velocities of the haloes, while satellite
galaxies are assumed to follow the spatial and velocity distributions of the
dark matter in the haloes. However, the HOD description itself allows the
freedom of varying the above assumptions, by introducing spatial bias and
velocity bias. For example, the recent modelling of small-scale redshift-space
clustering measurements using both the Sloan Digital Sky Survey (SDSS) Main
galaxy sample \citep{Guo15c} and SDSS-III Baryon Oscillation Spectroscopic
Survey \citep{Guo15a} shows that central galaxies have velocity offsets with
respect to the halo bulk velocities and the velocity distribution of
satellite galaxies generally differs from that of the dark matter. By
including such velocity bias factors, the HOD model is able to reproduce the
observed galaxy two-point correlation functions (2PCFs) in both projected and
redshift spaces remarkably well and to interpret successfully higher-order
statistics, like the three-point correlation functions \citep{Guo15b}.

The development of the high-resolution $N$-body simulations enables the
identification of the substructures within the dark matter haloes, i.e. the
subhaloes, which were distinct haloes before they fell into the current host
haloes \citep[see e.g.][]{Klypin16,Pujol14}. As in the literature, we refer to
virialized haloes that are not subhaloes of another halo as distinct haloes.
The subhaloes are believed to be the natural local environments for the
satellite galaxies in the host haloes. Due to their trackable merger
histories, the subhaloes provide a powerful way to study the galaxy evolution
once the connection between satellite galaxies and subhaloes is built. The
basic idea of the SHAM method is to assume a monotonic relation between
certain galaxy property and certain halo (including subhalo) property. For
example, the one-to-one correspondence between the galaxies and the dark
matter haloes (and subhaloes) can be made by ranking the galaxies in order of
their luminosity and populating the more massive haloes (and subhaloes) with
more luminous galaxies, i.e. the number density of galaxies above a
luminosity threshold is matched to that of haloes above a mass threshold,
establishing a link between galaxy luminosity and halo mass. In this way, the
galaxies relating to the host haloes are naturally central galaxies while
those in the subhaloes are satellite galaxies. In practice, the SHAM method
always includes a scatter in the galaxy-halo/subhalo relation, which has its
physical origin.

Accurately identifying and defining the subhaloes in the simulations should
take into account the effects of both the simulation resolution and baryon
physics \citep{Weinberg08}. While the resolution effect is less severe with
the emergence of more and more high-resolution simulations, the baryon
physics can still give rise to an important systematic effect for the SHAM
method. Compared to the stellar components of satellite galaxies that are
more gravitationally bound, the dark matter in subhaloes suffers more from
tidal heating and stripping. Galaxy properties are therefore more closely
connected to subhalo properties that are less affected by the tidal effects.
The original SHAM method is improved by relating the satellite galaxy
properties to the maximum circular velocity or the mass of subhaloes at the
epoch of accretion \citep[see e.g.][]{Conroy06,Vale06} or over the entire
merger history \citep[see e.g.][]{Moster10,Reddick13}. Such improvement is
shown to reproduce better the observed galaxy clustering measurements.
However, some effects are yet to be taken into account in the SHAM model. For
example, some subhaloes can be tidally destructed while the corresponding
satellite galaxies (stellar component) can still survive \citep[the so-called
orphan galaxies;][]{Wang06,Moster10}, and the usual SHAM model based on $N$-body
simulations would miss such a population.

The different halo and subhalo models have been studied extensively in the previous literature \citep[see e.g.][]{Yang12}. \cite{Yang09} used CLF method to explore the consequence of the stellar mass evolution of the satellite galaxies assuming the same stellar-halo mass relation (SHMR) for host haloes at present day and subhaloes at the time of accretion. They used the galaxy group catalogues \citep{Yang05c} constructed from SDSS DR4 to predict the stellar mass function of the satellite galaxies and emphasize the importance of including intracluster stars in the galaxy evolution. \cite{Neistein11a} studied the SHMR for central and satellite galaxies in the SHAM using a set of semi-analytical models (SAMs). They found that adopting the same SHMR for central and satellite galaxies cannot reproduce the clustering measurements in SAMs. \cite{Neistein11} further extended the SHAM models by allowing the stellar mass of the satellite galaxies to also depend on the host halo mass and concluded that the SHMR is not well constrained from the clustering measurements alone. \cite{Rodriguez-Puebla12} also found that different SHMRs for central and satellite galaxies are favoured by the observation by using the central and satellite stellar mass functions from the galaxy group catalogues. The SHAM technique is also examined in the smoothed particle hydrodynamics simulations by \cite{Simha12}, and it is found to overpopulate massive haloes because of severe stellar mass loss of some satellite galaxies. \cite{Reddick13} compared the connection between different halo properties and the galaxy stellar mass in the SHAM models. The scatter between galaxy stellar mass and halo property is constrained by the galaxy clustering measurements and the conditional stellar mass functions. They found that the model with the halo peak circular velocity provides the best agreement with the data.

The galaxy projected 2PCFs have been extensively used previously in constraining the models. However, the redshift-space clustering measurements have additional information about the galaxy velocity field and therefore can help distinguish different models. In this paper, we compare quantitatively the HOD and (extended) SHAM methods in modelling both the projected and redshift-space clustering of the volume-limited luminosity-threshold galaxy samples in the SDSS Data Release 7 (DR7). The galaxy-halo connections for the central and satellite galaxies are allowed to be different in the extended SHAM models. Unlike \cite{Rodriguez-Puebla12}, who apply SHAM separately to central and satellite stellar mass functions based on a group catalogue, we constrain all parameters of the extended SHAM models using the galaxy clustering measurements and the galaxy sample number densities.
In Section 2, we describe the
measurements of our galaxy samples and the modelling method. The subhalo
distributions in the high-resolution simulations are investigated in Section
3. We present the results of modelling the projected and redshift-space
clustering measurements in Sections 4 and 5, respectively. Finally, we
summarize our results and discuss the possible applications in Section 6.
Throughout the paper, we assume a spatially flat $\Lambda$ cold dark matter cosmology, with $\Omega_m=0.307$, $h=0.678$, and
$\sigma_8=0.823$, consistent with the constraints from Planck
\citep{Planck14}. The halo mass used in this paper is calculated based on the
given spherical overdensities of a viral structure \citep{Bryan98}.

\section{Measurements and Models}
In this paper, we use the galaxies in the New York University Value-Added
Galaxy Catalog \citep[NYU-VAGC;][]{Blanton05b} for the SDSS DR7
Main galaxy sample \citep{Abazajian09}. We further construct eight
volume-limited luminosity threshold samples, with absolute r-band Petrosian
magnitude $M_r$ varying from $-18$ to $-21.5$ with step size of $0.5$. We
refer the readers to \citet[][hereafter G15]{Guo15c} for more details.

The projected 2PCF $w_p(r_p)$ and
redshift-space 2PCF monopole ($\xi_0(s)$), quadrupole ($\xi_2(s)$) and
hexadecapole ($\xi_4(s)$) moments are measured for each sample, where $r_p$
and $s$ are the transverse and redshift-space separations of galaxy pairs,
respectively. The galaxy 2PCF measurements range from small scales of
$0.1\mpchi$ to intermediate scales of $25\mpchi$. The projected 2PCF $w_p(r_p)$ is measured by integrating the redshift-space 3D 2PCF to a maximum light-of-sight pair separation of $40\mpchi$ (also adopted in all the models). The covariance matrix for
each sample is estimated from jackknife resampling method
\citep{Zehavi11,Guo13}.

We follow the simulation-based model method laid out in \citet{Zheng16} to 
interpret the galaxy 2PCF measurements within the HOD and SHAM frameworks. 
It has been used in G15 and \citet{Guo15a}. 
With haloes identified in a high-resolution N-body simulation, this method tabulates all the necessary halo components in calculating galaxy 2PCFs, including one-halo pair distributions and two-halo 2PCFs from pairs composed of different combinations of central and satellite galaxies. With such tables and a specified description/parametrization of galaxy-halo relation (e.g. within the HOD and SHAM frameworks), galaxy 2PCFs are simply obtained by summing over different, pre-calculated table elements, weighted by the corresponding galaxy occupation statistics. With a given set of HOD (and SHAM) parameters, this method is equivalent to, but more efficient than, directly assigning galaxies to haloes (and subhaloes) in the simulation and measuring the corresponding model 2PCFs. Compared to  analytical models, it ensures high accuracy by using the halo information directly from the simulations and by calculating 2PCFs with exactly the same binning scheme as in the data. Finally, this method provides an efficient way to explore the parameter space for different models, which serves well our purpose in this paper.

We use the MultiDark simulation of
Planck cosmology (MDPL\footnote{The simulation is named as MDPL2 and publicly
available at
https://www.cosmosim.org/cms/simulations/multidark-project/mdpl2/};
\citealt{Klypin16}), with the cosmological parameters of $\Omega_m=0.307$,
$\Omega_b=0.048$, $h=0.678$, $n_s=0.96$, and $\sigma_8=0.823$. The simulation
has a volume of 1\,$h^{-3}$\,Gpc$^3$ (comoving) and the mass resolution is as
low as $1.51\times10^9\msun$. The simulation output at $z=0$ is adopted to
model all our luminosity threshold galaxy samples.
To see how simulation
resolution affects the subhalo population, we also investigate a smaller
simulation that has the same cosmological parameters as MDPL, but with a
volume of $0.4^3\,h^{-3}$\,Gpc$^3$, which is referred to as SMDPL
\citep{Klypin16}. This simulation was run with the same number of particles
($3840^3$) as in MDPL, so its mass resolution is $9.6\times10^7\msun$, about
15.6 times finer than MDPL. 

In both MDPL and SMDPL, the dark matter haloes and subhaloes are identified with the Rockstar
phase-space halo finder \citep{Behroozi13}, where the spherical haloes are
found from the density peaks in the phase space. The Rockstar code is
efficient and accurate to find the bound (sub)structures in the simulations
\citep{Onions12,Knebe13}. Note that different from G15, the unbound particles
are removed from our halo (and subhalo) catalogue. The halo (subhalo)
velocities are defined as the average particle velocity within the innermost
10\% of the halo (subhalo) radius, which is different from the definition of
centre-of-mass velocity (i.e. bulk velocity) of haloes in G15. The different
halo velocity definitions will affect the inferred galaxy velocity bias
parameters. This change of halo definition is to match those in the publicly
available Rockstar halo and subhalo catalogues. However, since we use the same
halo catalogues for the HOD and SHAM models, the comparison in this paper is
not affected by the definitions of haloes and halo properties. We consider
three sets of models to connect galaxies to the dark matter haloes in the
following sections. To avoid confusion, the host haloes and distinct haloes
mentioned hereafter refer to the haloes that are not subhaloes of any other
dark matter haloes.

\subsection{The HOD Model}
For a sample of galaxies above a given luminosity threshold, the HOD model
includes five parameters for describing the average number $N$ of galaxies in
distinct haloes of mass $M_{\rm h}$ \citep*{Zheng07}
%%%%%%%%%%%%%%%%%%%%%%%%%%%%%%%%%%%%%%%%%%%%%%%%%%
\begin{eqnarray}
\langle N(M_{\rm h})\rangle&=&\langle N_{\rm cen}(M_{\rm h})\rangle+\langle N_{\rm sat}(M_{\rm h})\rangle,\\
\langle N_{\rm cen}(M_{\rm h})\rangle&=&\frac{1}{2}\left[1+{\rm erf}\left(\frac{\log M_{\rm h}-
\log M_{\rm min}}{\sigma_{\log M_{\rm h}}}\right)\right], \label{eqn:Ncen}\\
\langle N_{\rm sat}(M_{\rm h})\rangle&=&\langle N_{\rm cen}(M_{\rm h})\rangle\left(\frac{M_{\rm h}-M_0}
{M_1^\prime}\right)^\alpha,
\label{eqn:Nsat}
\end{eqnarray}
%%%%%%%%%%%%%%%%%%%%%%%%%%%%%%%%%%%%%%%%%%%%%%%%%%
where the two central galaxy parameters $M_{\rm min}$ and $\sigma_{\log
M_{\rm h}}$ describe the characteristic minimum mass of haloes that host the
sample of galaxies ($\langle N_{\rm cen}(M_{\rm min})\rangle=0.5$) and the
characteristic width of the transition mass range for haloes hosting zero to
one galaxy. The three parameters for the satellite galaxies are the cutoff
mass scale $M_0$, the normalization mass scale $M_1^\prime$ and the power-law
slope $\alpha$ at the high-mass end. In this paper, we fix $\alpha\equiv1$ in
order to match the slope of the subhalo occupation function in massive haloes
and to reduce the degrees of freedom (dof) to match that in the SHAM model
(see below). In the following sections, we also compare two useful derived
parameters, the characteristic mass $M_1$ of haloes hosting on average one
satellite galaxy and the inferred satellite fraction $f_{\rm sat}$
(defined as the fraction of the satellite galaxies in the sample).

We note that to compute the mean number of intra-halo central-satellite pairs
in the model, the occupation numbers of central and satellite galaxies are
assumed to be independent of each other. That is, we have $\langle N_{\rm
cen}N_{\rm sat}\rangle=\langle N_{\rm cen}\rangle\langle N_{\rm sat}\rangle$.
Changing the assumption of the dependence between the central and satellite occupations only has minimal effects on the HOD parameters, as discussed in Fig.~10 of \cite{Guo15a}. Compared to the case of having satellites only in haloes with central galaxies for a given galaxy sample, we now can populate satellites in some low mass haloes without central galaxies. As a consequence, the best-fitting $\alpha$ will decrease and the central galaxy velocity bias will slightly shift to lower values, while other HOD parameters only change by about 0.1\%.

In our fiducial model, the central galaxies are assigned the positions and
velocities of the distinct haloes, while the random dark matter particles in
the haloes are selected to represent the satellite galaxies. As in G15, we
introduce an additional central galaxy velocity bias parameter $\alpha_c$ in
the HOD model to allow the central galaxy velocity to differ from that of the
halo velocity, with a velocity dispersion equal to $\alpha_c$ times 
the dark matter particle velocity dispersion $\sigma_v$ in the haloes. We
also include the satellite velocity bias parameter $\alpha_s$. The relative
velocity of a satellite galaxy to the halo centre is scaled by the satellite
velocity bias $\alpha_s$ to take into account the possible velocity
differences between the dark matter particles and the satellite galaxies. In
the frame of a single halo, the satellite galaxy velocity bias is the same as
the ratio between the velocity dispersions of the satellite galaxies
($\sigma_{\rm sat}$) and the dark matter particles within the haloes, i.e.
$\alpha_s=\sigma_{\rm sat}/\sigma_v$. We refer the readers to G15
for more details. In total, we have six free parameters in the HOD model, four
for the mean occupation function ($M_{\rm min}$, $\sigma_{\log M_h}$, $M_0$, 
and $M_1^\prime$) and two for the velocity bias ($\alpha_c$ and $\alpha_s$).

We apply a Markov Chain Monte Carlo (MCMC) method to explore the probability
distribution of the model parameters. The likelihood surface is determined by
$\chi^2$, contributed by the projected 2PCF $w_p$, the redshift-space
multipoles $\xi_0$, $\xi_2$ and $\xi_4$, and the observed galaxy number
density $n_g$,
%%%%%%%%%%%%%%%%%%%%%%%%%%%%%%%%%%%%%%%%%%%%%%%%%%
\begin{equation}
\chi^2= \bmath{(\xi-\xi^*)^T C^{-1} (\xi-\xi^*)}
       +\frac{(n_g-n_g^*)^2}{\sigma_{n_g}^2}, \label{eq:chi2}
\end{equation}
%%%%%%%%%%%%%%%%%%%%%%%%%%%%%%%%%%%%%%%%%%%%%%%%%%
where $\bmath{C}$ is the full error covariance matrix and the data vector
$\bmath{\xi} = [\bmath{w_p},\bmath{\xi_0},\bmath{\xi_2},\bmath{\xi_4}]$. The
quantity with (without) a superscript `$*$' is the one from the measurement
(model). To take into account the finite volume of the simulations our model
is based on, we also apply a volume correction of $1+V_{\rm obs}/V_{\rm sim}$
to the covariance matrix \citep[][]{Zheng16}, where $V_{\rm obs}$ and $V_{\rm
sim}$ are the volumes for the observed galaxy sample and the simulation,
respectively. For each sample and each model, we perform MCMC runs with length of two million to explore the parameter space and to choose the set of best-fitting parameters. For the chain, at each step of the random walk,
a set of trial HOD parameters are generated. Covariances among parameters are taken into account when proposing the trial move in order to improve the efficiency of the chain. The probability of keeping the 
trial HOD parameters depends on the difference $\Delta\chi^2=\chi^2_{\rm new}-\chi^2_{\rm old}$ between the old and new (trial) sets of parameters, i.e. 1 for 
$\Delta\chi^2\leq0$ and $\exp(-\Delta\chi^2/2)$ for $\Delta\chi^2>0$.

\subsection{The SHAM Models}\label{subsec:sham}
The simplest SHAM model usually assumes a monotonic relation between the
galaxy luminosity (or stellar mass) and a given halo property (e.g. halo
mass), by assigning more luminous galaxies to more massive haloes. The galaxy
luminosity function is then preserved by matching the number density of the
galaxy sample to that of the haloes \citep[see e.g.][]{Conroy06}. Since such
an assignment is only based on the halo property (e.g. halo mass), the
distinct halo and subhalo in the simulations are not distinguished between
each other. The relation between the galaxies and the haloes (including both
distinct haloes and subhaloes) is completely determined by the number density
distribution (e.g. luminosity function) of the galaxy sample. Thus, there is
no free parameter in such models. A more flexible SHAM model is typically
introduced to allow a scatter between e.g. the galaxy luminosity and the halo
mass. Such a scatter is necessary especially when modelling the clustering of
the luminous galaxies \citep[see e.g.][]{Reddick13}.

There are a few popular SHAM models that connect the galaxy luminosity to the
different halo properties. In this paper, we only consider the following
three SHAM models using different halo properties.

(1) $M_{\rm acc}$. For a distinct halo, it is the current halo mass, while
for a subhalo, it is the mass at the last epoch when the subhalo was a
distinct halo (before accreted to another halo).

(2) $V_{\rm acc}$. For a distinct halo, it is the current maximum circular
velocity, while for a subhalo, it is the maximum circular velocity at the
last epoch of being a distinct halo (before accreted to another halo).

(3) $V_{\rm peak}$. For both distinct haloes and subhaloes, it is the peak
circular velocity over the entire merger history.

The properties $M_{\rm acc}$ and $V_{\rm acc}$ are commonly used in the SHAM
models because they are closely related to the halo merger history, while
recent results suggest that choosing $V_{\rm peak}$ in the model leads to
better agreement with the data \citep[e.g.][]{Moster10}. The $V_{\rm peak}$ of a distinct halo or subhalo is usually significantly larger than $V_{\rm acc}$, because the peak circular velocity is generally achieved earlier in time than the accretion. The tidal heating and stripping will later reduce the circular velocity of a subhalo even before the accretion \citep[see e.g. Fig.~1 of][]{Chaves-Montero15}. \cite{Reddick13} compared different SHAM models and found that $V_{\rm peak}$ is more closely related to the galaxy stellar mass, while $M_{\rm peak}$ (the maximum mass that a halo or subhalo has ever had in its merger history) is generally not successful in reproducing the clustering measurements. So we do not consider the $M_{\rm peak}$ case in our SHAM models. We will
investigate these three models in the following sections.

In implementing the SHAM models we allow a scatter between the galaxy
property (here luminosity) and the adopted halo property. To facilitate the
comparison with the HOD model, the scatter is parametrized in a way of using
the functional form of Eq.~\ref{eqn:Ncen} to assign galaxies to haloes. As an
example of choosing $M_{\rm acc}$ as the halo property, the probability of a
distinct halo or subhalo having a galaxy in a given luminosity-threshold
sample is
\begin{equation}
P(M_{\rm acc}) =\frac{1}{2}\left[1+{\rm erf}\left(\frac{\log M_{\rm acc}-
\log M_{\rm min,acc}}{\sigma_{\log M_{\rm acc}}}\right)\right] \label{eq:macc}
\end{equation}
The scatter between galaxy property and halo property is encoded in the
parameter $\sigma_{\log M_{\rm acc}}$ \citep[][]{Zheng07}, which is the only
free parameter in Eq.~\ref{eq:macc}. The characteristic mass scale $M_{\rm
min,acc}$ can then be determined by matching the sample number density. For
other two halo properties, we only need to replace the mass in
Eq.~\ref{eq:macc} to the corresponding terms for $V_{\rm acc}$ and $V_{\rm
peak}$. Note that the SHAM model we use here is more flexible than the
commonly adopted one. The usual SHAM model assumes one scatter parameter and
performs the abundance matching for galaxies in the full range of observed
luminosity. Here we model a series of luminosity-threshold samples, and each
has its own scatter parameter. We are effectively allowing the scatter
between the galaxy luminosity and the halo property to vary with the halo
property.

In the SHAM model we use, a further improvement is related to the
determination of the scatter parameter. We do not simply assign a scatter
parameter for a given luminosity-threshold sample. The final $\sigma_{\log
M_{\rm acc}}$ used in each luminosity-threshold sample is determined from the
model with the best-fitting $\chi^2$ to the galaxy projected 2PCFs. We
emphasize that even though the scatter parameter we introduce here is
formally expressed in terms of the halo property (mass or circular velocity),
it is originally derived from the scatter in the (lognormal) galaxy
luminosity distribution at a fixed halo mass or circular velocity \citep[see
Eq.~4 in][]{Zheng07}. The meaning of $\sigma_{\log M_{\rm acc}}$ is not the
scatter on the halo mass at a fixed galaxy luminosity, but rather the width
of the cutoff profile. We can conveniently convert $\sigma_{\log M_{\rm
acc}}$ to the scatter on the galaxy luminosity $\sigma_{\log L}$ at fixed
halo mass using the local slope of the $L$--$M_{\rm acc}$ relation at the
threshold luminosity, as will be shown in the following sections.

For central galaxy occupation distribution in the $M_{\rm acc}$ model, we can
directly compare $M_{\rm min,acc}$ to $M_{\rm min}$ in the HOD model, because
they both refer to the typical cutoff mass of the distinct haloes that host
the galaxies in the sample of interest. For satellite galaxies in subhaloes
of $M_{\rm acc}$ at the time of accretion, with the simulations we can
conveniently convert $P(M_{\rm acc})$ in Eq.~\ref{eq:macc} to the satellite
mean occupation function $\langle N_{\rm sat}(M_{\rm h})\rangle$ in host 
haloes of mass $M_{\rm h}$. From the average occupation number $\langle N_{\rm
sub}(M_{\rm acc}|M_{\rm h})\rangle$ of subhaloes with mass $M_{\rm acc}$ in
each host halo with mass $M_{\rm h}$, we have
\begin{equation}
\langle N_{\rm sat}(M_{\rm h})\rangle = \sum_{M_{\rm acc}}
P(M_{\rm acc})\langle N_{\rm sub}(M_{\rm acc}|M_{\rm h})\rangle.
\label{eq:satmacc}
\end{equation}
For the cases of $V_{\rm acc}$ and $V_{\rm peak}$ models, the mean satellite
function can be computed similarly by replacing the mass in
Eq.~\ref{eq:satmacc} to the corresponding velocity variable.

Overall, the SHAM model we use here is more flexible, compared to the
traditional one. We allow the scatter to depend on the halo property, and
determine it by fitting the projected 2PCF. The number density of the galaxy
sample is ensured to be matched by tuning the characteristic halo mass scale
$M_{\rm min,acc}$. In what follows, we further extend or generalize the SHAM
model to make it even more flexible, with the relevant parameters determined
by both the galaxy abundance and the galaxy clustering (in projected and
redshift spaces).

\subsection{A Subhalo Clustering and Abundance Matching Model}
The galaxy luminosity (or halo mass/property) dependent scatter extends the
SHAM models. However, as will be shown below, this extension is still not
capable of satisfactorily interpreting the observed galaxy 2PCFs. We
therefore add further flexibilities to the SHAM model and make it a well
parametrized model to fit both the galaxy abundance and clustering, which
can be referred to as subhalo clustering and abundance matching (SCAM) model.

For a given luminosity-threshold galaxy sample, we construct the SCAM model
by allowing the mass scale $M_{\rm min,acc}$ and scatter parameter
$\sigma_{\log M_{\rm acc}}$ in Eq.~\ref{eq:macc} to be different for the
distinct haloes (central galaxies) and subhaloes (satellites). That is, we
now have probabilities $P_{\rm cen}(M_{\rm acc})$ and $P_{\rm sat}(M_{\rm
acc})$. The extensions for the case of $V_{\rm acc}$ and $V_{\rm peak}$ are
similar. Once a halo property is chosen to use, we have four parameters for
the central and satellite mean occupation functions. Such separate
parametrizations for the central and satellite components in the SCAM model
are supported by the recent findings of the differences between the central
and satellite galaxies in the SHAM models
\citep{Rodriguez-Puebla12,Watson13}.

To model the redshift-space 2PCFs with the SCAM model, the treatment of the
central galaxies is the same as in the HOD model and a central galaxy
velocity bias parameter $\alpha_c$ is introduced. Since the subhaloes are
selected to host satellite galaxies, we also apply a satellite galaxy
velocity bias by scaling the velocity of a subhalo relative to its host halo
with a factor of $\alpha_s$. So in total we have six free parameters for the
redshift-space modelling with the SCAM model. As with the HOD model, the
parameter space is explored with the MCMC method with the likelihood
determined by the 2PCFs and the galaxy number density (Eq.~\ref{eq:chi2}).

\section{Particle and Subhalo Distributions in Simulations}\label{sec:sub}
Before we apply the HOD/SHAM/SCAM models to model the clustering
measurements, it is important to understand the particle and subhalo
distributions in the simulations. As subhaloes are related to satellites in
SHAM/SCAM, the HOD model in this paper connects satellites to dark matter
particles. Any difference seen in the particle and subhalo distributions will
be useful for us to understand the modelling results. 

\begin{figure*}
\includegraphics[width=0.8\textwidth]{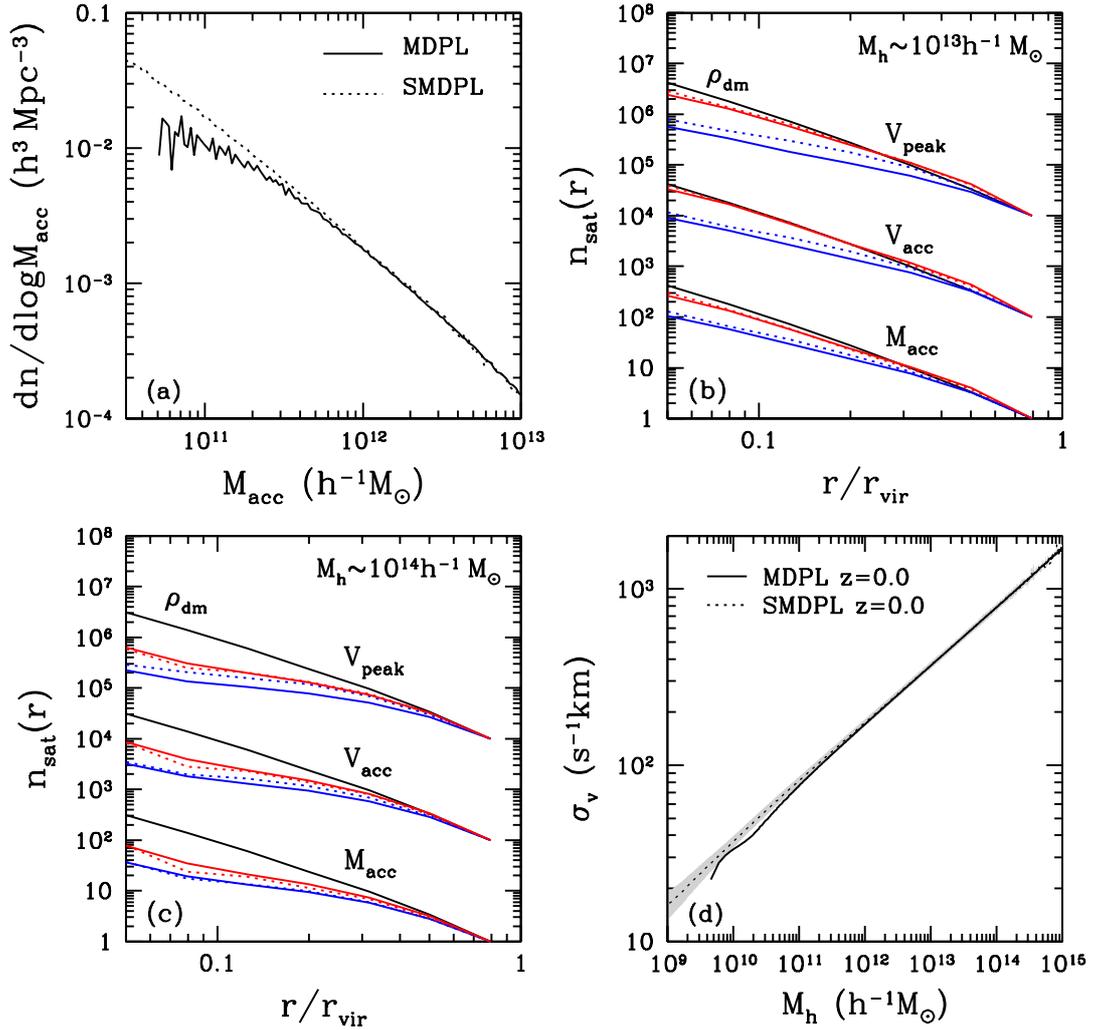}
\caption{Comparisons of the subhalo distributions between the MDPL and SMDPL
simulations. In each panel, solid and dotted curves are from the MDPL and SMDPL
simulations, respectively.
Panel (a): subhalo mass functions.
Panel (b): subhalo spatial distribution profile in the host haloes of
$M_{\rm h}{\sim}10^{13}\msun$. The red and blue curves are for subhaloes
selected using different mass or velocity thresholds. For the $M_{\rm acc}$
model, the red and blue curves are for $M_{\rm acc}>10^{12}$ and
$>10^{11.5}\msun$, respectively. For the $V_{\rm acc}$ and $V_{\rm peak}$
models, the red and blue curves are for $V_{\rm acc}$ (or $V_{\rm peak}$)
larger than $10^{2.3}$ and $10^{2.1}\kms$, respectively. For each model,
the profiles are normalized to be the same at the host halo virial radius and
the curves are separated for different models for clarity. The black solid
lines are the density profiles for the dark matter particles in each case.
Panel (c): similar to panel (b), but for the host haloes of $M_{\rm
h}{\sim}10^{14}\msun$. Panel (d): 3D dark matter velocity dispersion in
distinct haloes of different mass $M_{\rm h}$. The shaded area shows the
scatter around the velocity dispersion measurements in SMDPL.}
\label{fig:profile}
\end{figure*}
We show in Fig.~\ref{fig:profile} the detailed comparisons between the
subhalo distributions in the MDPL and SMDPL simulations. Panel (a) shows the
subhalo mass functions in the two simulations. The simulation resolution does
affect the identification of the subhaloes in the two simulations. But for
subhaloes of $M_{\rm acc}>2.8\times10^{11}\msun$, the subhaloes in MDPL are
about 90\% complete, compared to that of the SMDPL. In terms of circular
velocities, subhaloes are 90\% complete in MDPL for $V_{\rm acc}>176\kms$ and
$V_{\rm peak}>184\kms$, respectively.

\begin{figure*}
\includegraphics[width=1.0\textwidth]{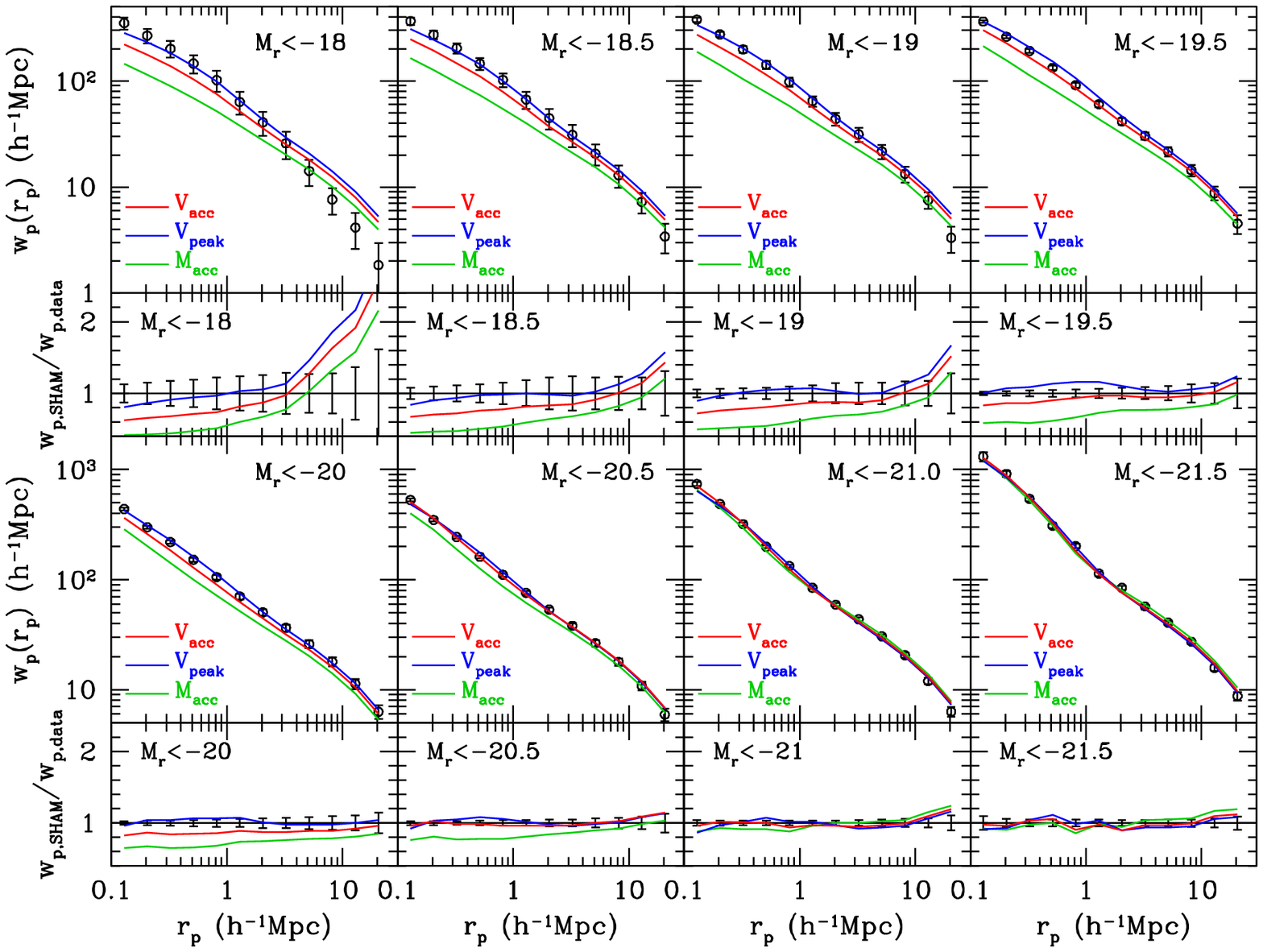}
\caption{Best-fitting models for the projected 2PCF $w_p(r_p)$ using the
different SHAM models with scatters.
The measurements for volume-limited samples in SDSS DR7 Main galaxies are
shown as the circles with error bars. The different SHAM models are shown as the
different colour lines as labelled. The ratios between the SHAM models and
the measurements are shown in the bottom part of each panel, with the error
bars from the measurements.} \label{fig:wpsham}
\end{figure*}
As will be shown in the following sections, many faint satellite galaxies in
the SHAM/SCAM model are predicted to reside in subhaloes of mass $M_{\rm
acc}$ around $10^{11}\msun$. The corresponding subhaloes identified in MDPL
simulation suffer from the resolution effect, so for the SHAM/SCAM method we
will model the faint galaxy samples of $M_r<-18$, $-18.5$, $-19$, and $-19.5$
using the SMDPL simulation instead and model the more luminous samples using
the MDPL simulation. The volume $V_{\rm sim}$ of the SMDPL is much larger than 
the survey volume $V_{\rm obs}$ of these faint samples (G15), so the volume 
correction (the $1+V_{\rm obs}/V_{\rm sim}$ factor) to the covariance matrix 
\citep{Zheng16} is not significant. For the HOD model, since we
are randomly selecting the dark matter particles to represent the satellite
galaxies, the resolution of the MDPL simulation is high enough to model all
the luminosity threshold samples. So we do not use the SMDPL for the HOD
models. We have verified that using SMDPL for modelling the faint galaxy
samples with the HOD method produces the same results as using the MDPL
simulation. This is consistent with the fact that the mass functions for the distinct haloes in MDPL and SMDPL agree down to haloes of about $5\times10^{10}\msun$ \citep{Rodriguez-Puebla16}.

Panels (b) and (c) display the number density profiles of subhaloes in host
haloes around $M_{\rm h}=10^{13}$ and $10^{14}\msun$ as a function of subhalo
properties ($M_{\rm acc}$, $V_{\rm acc}$, and $V_{\rm peak}$, as labelled).
For each subhalo property, the density profiles are normalized to be the same
at the host halo virial radius and offsets are added for the curves of
different subhalo properties for clarity. In each set of curves, the black
solid line is the density profile of the dark matter particles. The solid
lines are for the subhalo density profiles in MDPL, while the dotted lines
are for those in the SMDPL. The red and blue curves are for subhaloes
selected using different mass or velocity thresholds. For the $M_{\rm acc}$
model, the red and blue curves are for $M_{\rm acc}>10^{12}$ and
$>10^{11.5}\msun$, respectively. For the $V_{\rm acc}$ ($V_{\rm peak}$)
model, the red and blue curves are for $V_{\rm acc}$ ($V_{\rm peak}$) larger
than $10^{2.3}$ and $10^{2.1}\kms$, respectively. In general, the density
profile of the subhaloes is shallower than that of the dark matter \citep[see
e.g.][]{Gao04,Pujol14}. But as the mass ratio $M_{\rm acc}/M_{\rm h}$ (or
velocity ratio) increases, the subhalo density profile is approaching that of
the dark matter. More importantly, such a trend is not affected by the mass
resolution of the simulations, which indicates that the scarce of subhaloes
in the inner regions of the host haloes is most likely caused by the strong
tidal stripping effect \citep[see e.g.][]{Springel08}. Since the stellar
components of satellite galaxies are more tightly bound, they can still
survive to be observed as satellites even if the corresponding subhaloes lose
their identities from tidal destruction. The possibly different distribution
profiles between subhaloes and satellite galaxies will then be an important
factor to consider when interpreting the clustering modelling results with
both the HOD and SHAM/SCAM models.

Panel (d) shows the 3D dark matter velocity dispersions $\sigma_v$ as a
function of the host halo mass $M_{\rm h}$. The two simulations show very
good agreement with each other. For distinct haloes with mass $M_{\rm
h}>10^{11}\msun$, the velocity dispersion measurements are not significantly
affected by the simulation resolutions.

\begin{figure}
\includegraphics[width=0.5\textwidth]{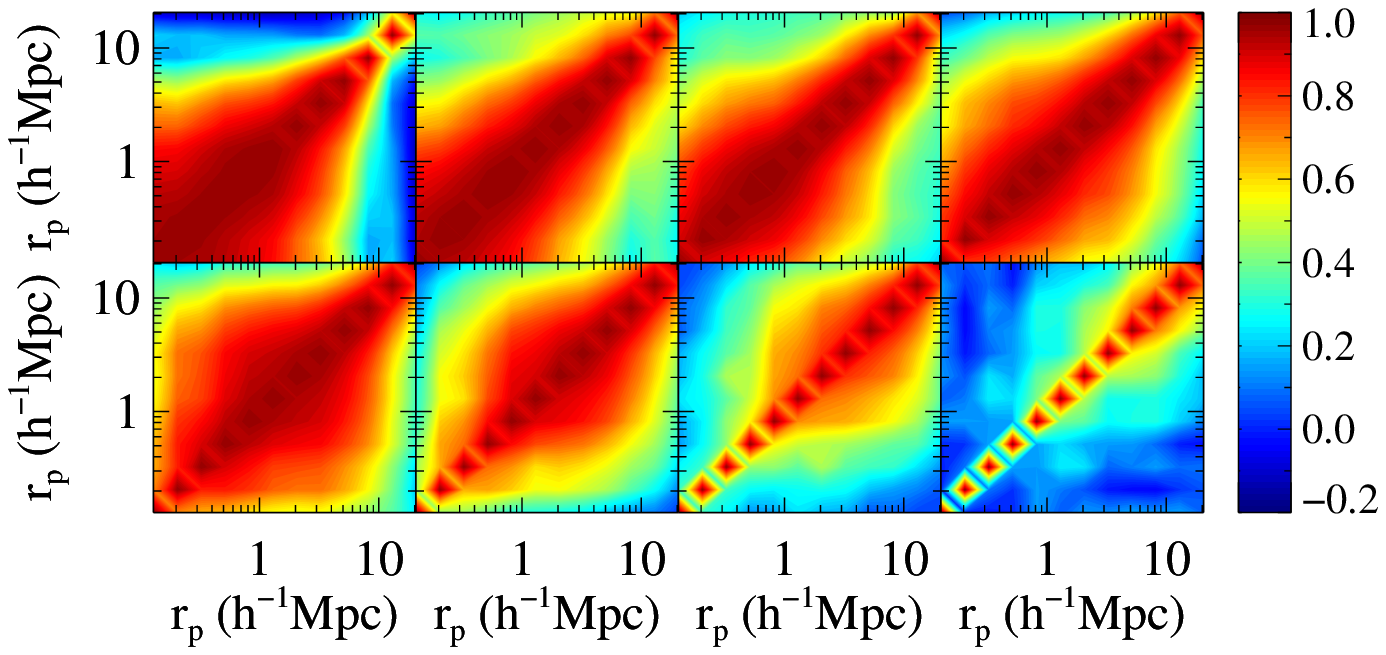}
\caption{Normalized covariance matrices for the corresponding 2PCF measurements shown in
Fig.~\ref{fig:wpsham}. From left to right and top to bottom, the covariance
matrices are for the luminosity threshold samples from $M_r<-18$ to $M_r<-21.5$.
}
\label{fig:cov}
\end{figure}
\begin{figure}
\includegraphics[width=0.42\textwidth]{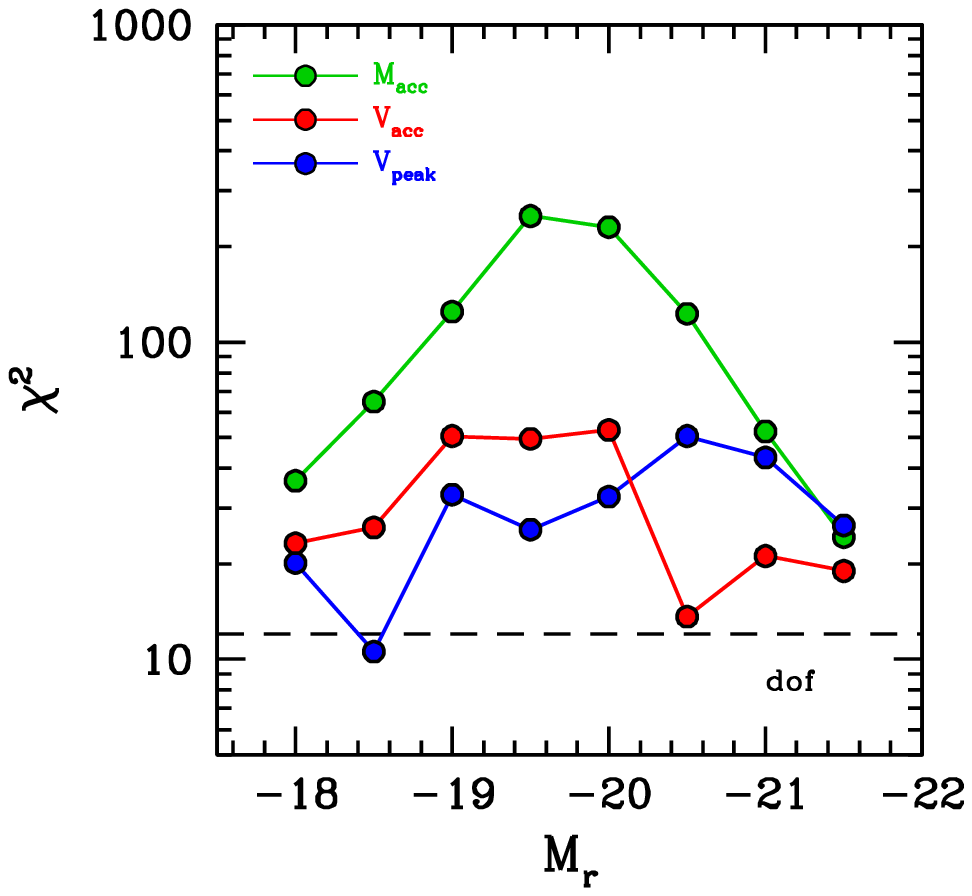}
\caption{Best-fitting $\chi^2$ of the different SHAM models from $w_p$--only
data for the
different luminosity threshold samples. The number of dof
of the models is shown as the horizontal dashed line.} \label{fig:chisham}
\end{figure}
Since we have the 3D velocity for each subhalo in the simulations, an
interesting question is the velocity bias of the subhaloes with respect to
the dark matter velocity distribution. We measure the velocity dispersions
$\sigma_{\rm sub}$ for subhaloes of different masses in different host
haloes, and estimate the average subhalo velocity bias $\alpha_{\rm sub}$
through the following equation,
\begin{equation}
\langle\alpha_{\rm sub}\rangle=\sqrt{\langle\sigma_{\rm sub}^2/\sigma_v^2\rangle},
\end{equation}
which is an unbiased estimate of the subhalo velocity bias even for a small
number of subhaloes in each host halo. The subhalo velocity dispersion
$\sigma_{\rm sub}$ in each halo is calculated by
\begin{equation}
\sigma_{\rm sub}^2=\frac{1}{N}\sum_{i=1}^N\bmath{\|v_{\rm sub}-v_{\rm h}\|}^2,\label{eq:sigsub}
\end{equation}
where $\bmath{v_{\rm sub}}$ and $\bmath{v_{\rm h}}$ are the 3D velocities of
the subhalo and the corresponding host halo, respectively, and $N$ is the
number of subhaloes of interest in each halo. Note that our definition of
subhalo velocity dispersion is different from that of \cite{Wu13b}, who used
the mean velocity of all the subhaloes in the host halo instead of
$\bmath{v_{\rm h}}$ in Eq.~\ref{eq:sigsub}. That is, we include the
dispersion in the offset between the mean velocity of subhaloes and the halo
velocity. Also, the subhalo velocity bias in \cite{Wu13b} is estimated
through $\langle\sigma_{\rm sub}/\sigma_v\rangle$, which is a biased
estimator of the velocity bias and needs corrections for small $N$. This can
be seen by considering a 1D velocity distribution with zero mean: while
$\sqrt{\langle v^2\rangle}$ gives the dispersion $\sigma$, in general
$\langle |v|\rangle$ (a.k.a. mean absolute deviation) does not. The reason
that we choose $\bmath{v_{\rm h}}$ as the reference velocity is to match the
way we define the satellite galaxy velocity bias in the HOD model. We measure
the subhalo velocity bias $\alpha_{\rm sub}$ for subhaloes with masses
$M_{\rm acc}>10^{11}\msun$ in haloes of different $M_{\rm h}$ in both
simulations. The measured $\alpha_{\rm sub}$ varies from 1.02 to 1.11 for
$M_{\rm acc}$ in the range of $10^{11}$--$10^{13}\msun$. The lower mass
subhaloes have slightly larger values of $\alpha_{\rm sub}$. This trend of
$\alpha_{\rm sub}$ with the subhalo mass is less significant than that 
in Fig.~1 of
\cite{Wu13b}. We find that even for the most massive subhaloes in their host
haloes, the value of $\alpha_{\rm sub}$ is still around $1$, which is much
larger than the value of about $0.8$ inferred from \cite{Wu13b} (We recover
the same values of $\alpha_{\rm sub}$ as in their Fig.~1 when switching to
their estimator). Note that the haloes and subhaloes in \cite{Wu13b} are also
identified using the Rockstar code. The above difference is mainly caused by
the biased estimator they use, with a small contribution from our choosing
$\bmath{v_{\rm h}}$ in evaluating the velocity dispersion.

As shown in G15, the satellite galaxy velocity bias $\alpha_s$ from HOD
modelling the redshift-space clustering of our sample is generally smaller
than 1, with a typical value of 0.8. Therefore, the difference between
$\alpha_s$ and $\alpha_{\rm sub}$ indicates the necessity of including
satellite velocity bias in the subhalo models when modelling the
redshift-space clustering using SHAM/SCAM.

\section{Modelling the Projected 2PCFs}\label{sec:wp}
In the following sections, we will consider the modelling of the projected 2PCF only ($w_p$), as well as the modelling of both the projected and redshift-space 2PCFs ($w_p+\xi_{0,2,4}$). To guide the readers, we list all the measurements and models used in the following sections in Table~\ref{tab:sec}.  When only the $w_p$ is used in constraining models, the contribution to $\chi^2$ from clustering will only include that from $w_p$ in Eq.~\ref{eq:chi2}, i.e. $\bmath{\xi}=\bmath{w_p}$.

We first consider the modelling of the projected 2PCF $w_p(r_p)$ only, which
is commonly used in constraining the HOD and SHAM parameters. In the
modelling of $w_p$, we do not include the velocity bias parameters, because
the projected 2PCF is integrated over the line of sight and hence 
relatively insensitive to the galaxy velocities.

\begin{table*}
\caption{Measurements used in the fits with different models} \label{tab:sec}
\begin{tabular}{lllll}
\hline
Measurements &  Models &  Number of Free Parameters & Section & Comments\\
\hline
$w_p$  & SHAM      &  1 & \S4.1 & $n_g$ exactly matched\\
$w_p+n_g$  & SCAM/HOD  &  4 & \S4.2 & \\
$w_p+\xi_{0,2,4}+n_g$  &  SCAM/HOD & 6 & \S5 & SHAM results also presented\\
\hline
\end{tabular}
\end{table*}

\subsection{Results from the SHAM Models}
\begin{figure*}
\includegraphics[width=0.8\textwidth]{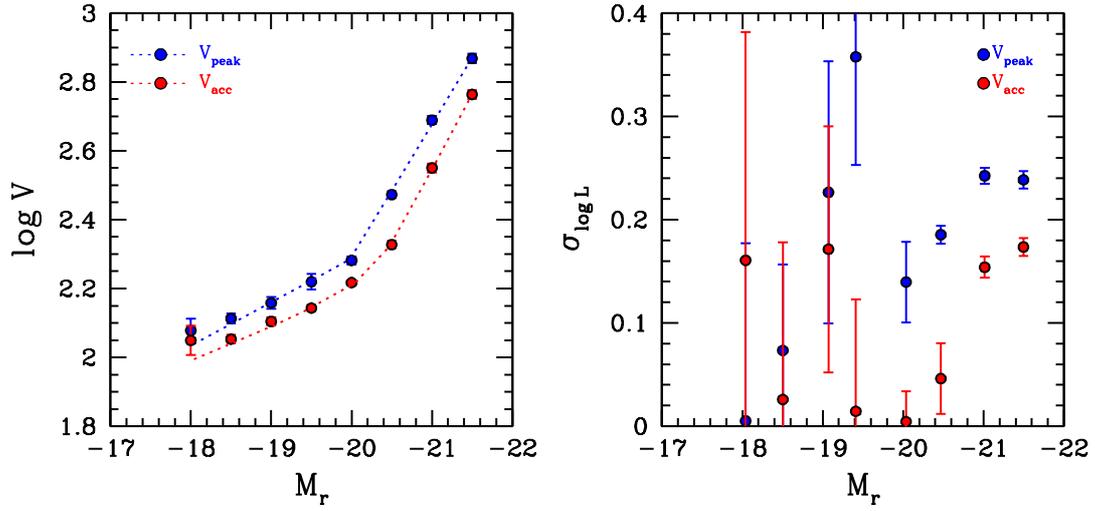}
\caption{Comparisons of the model parameters for the $V_{\rm acc}$ and
$V_{\rm peak}$ models from fitting the $w_p$--only data.
The left-hand panel shows the characteristic cutoff
circular velocity as a function of sample luminosity threshold
for the two models. The right-hand panel shows the corresponding
scatters in galaxy luminosity in haloes with circular velocities around the
cutoff velocity (see the text).} \label{fig:siglogl}
\end{figure*}
\begin{figure*}
\includegraphics[width=1.0\textwidth]{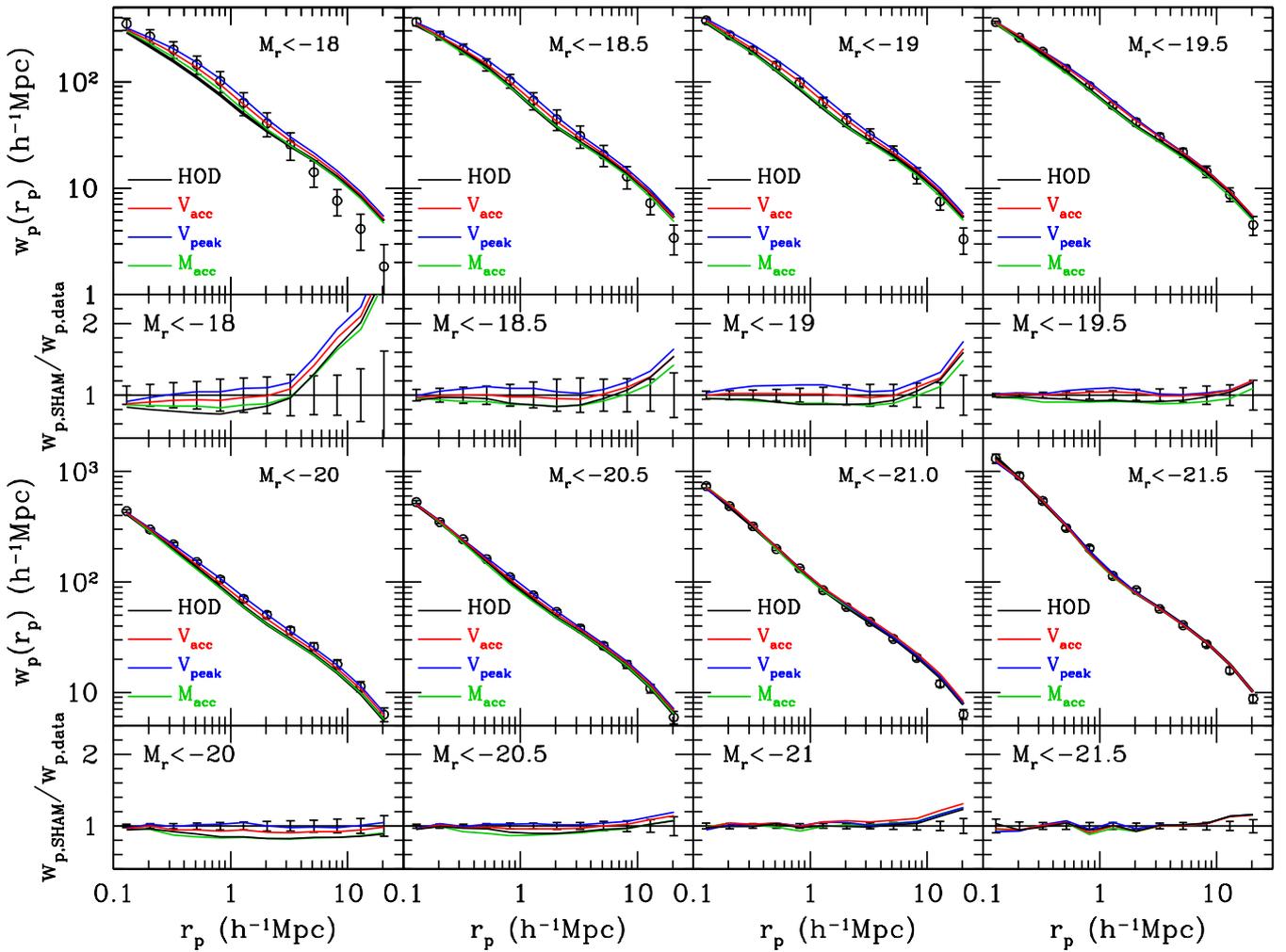}
\caption{Similar to Fig.~\ref{fig:wpsham}, but for the SCAM models. The
best-fitting HOD models are also included, shown as the black lines.} \label{fig:wpshamc}
\end{figure*}
\begin{figure*}
\includegraphics[width=0.8\textwidth]{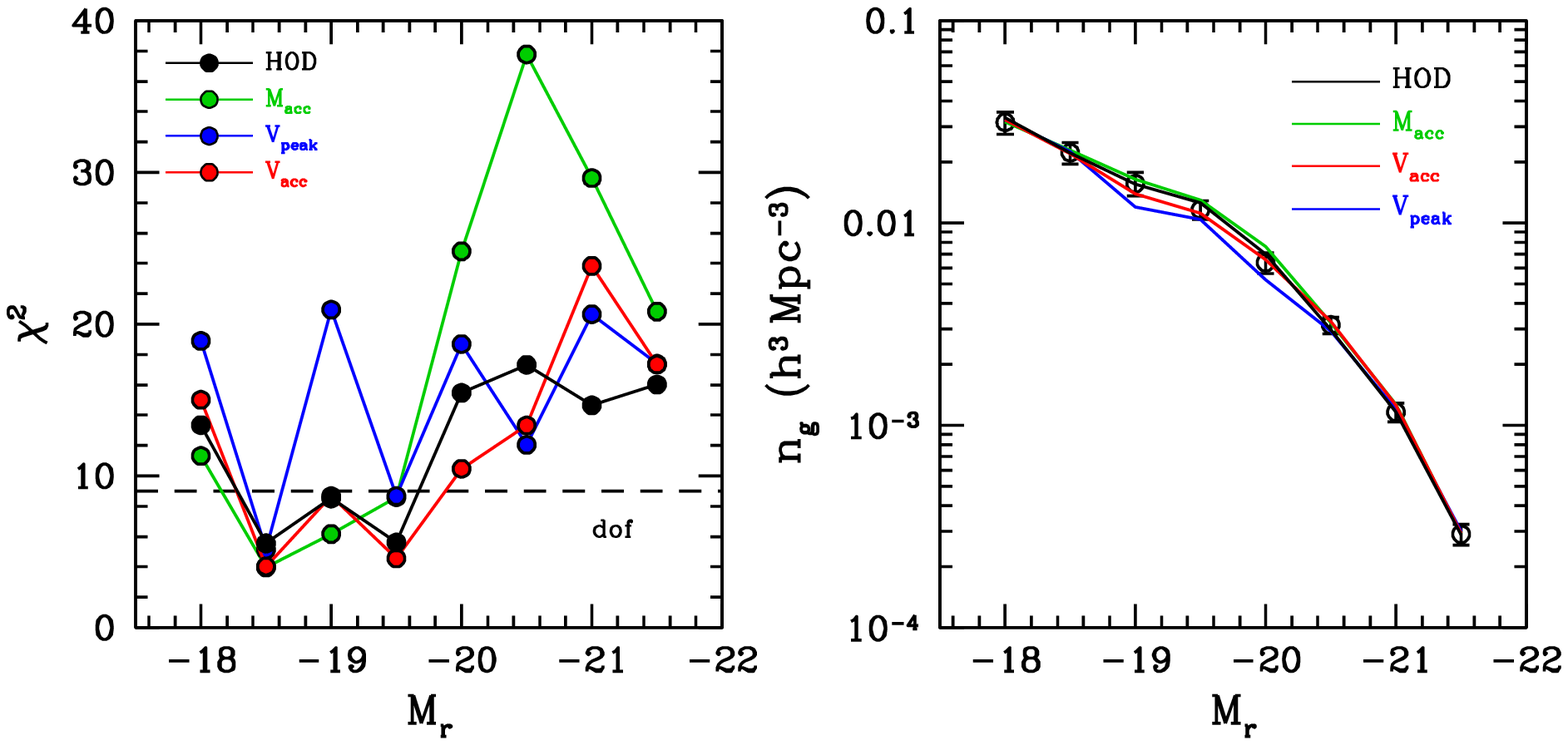}
\caption{Left: best-fitting $\chi^2$ of the different models from fitting
$w_p$--only data for the
different luminosity threshold samples. The number of dof of the
models is shown as the horizontal dashed line. Right: comparison between the
galaxy number densities (curves) from the best-fitting models and the
measured ones (circles).} \label{fig:chishamc}
\end{figure*}

We first compare the modelling results from the three SHAM models (based on $M_{\rm acc}$, $V_{\rm acc}$, and $V_{\rm peak}$, respectively) including
scatters as described in \S\ref{subsec:sham}. Fig.~\ref{fig:wpsham} shows the
best-fitting SHAM models to $w_p(r_p)$ for the eight volume-limited
luminosity threshold samples in SDSS DR7. The different SHAM models are shown
as the different colour lines. Overall, the $V_{\rm peak}$ model seems to
provide the best descriptions for all the galaxy samples, consistent with the
conclusions of \citet{Reddick13}. The $M_{\rm acc}$ and $V_{\rm acc}$ models
significantly underestimate the small-scale clustering for faint galaxies of
threshold luminosity $M_r$ fainter than $-20.5$. This can be attributed to
the shallower subhalo distribution profiles (Fig.~\ref{fig:profile}). The
$V_{\rm peak}$ model provides better fittings to the data, because the values
of $V_{\rm peak}$ for subhaloes are usually much larger than $V_{\rm acc}$.
We note that in Fig.~\ref{fig:profile} the red and blue curves for $V_{\rm
acc}$ and $V_{\rm peak}$ are selected using the same thresholds. For the same
galaxy sample, the thresholds of $V_{\rm acc}$ and $V_{\rm peak}$ would be
different, and the density profiles for the subhaloes selected using the
best-fitting $V_{\rm peak}$ model is closer to the dark matter distribution
than using the best-fitting $V_{\rm acc}$ model.

\begin{figure*}
\includegraphics[width=0.9\textwidth]{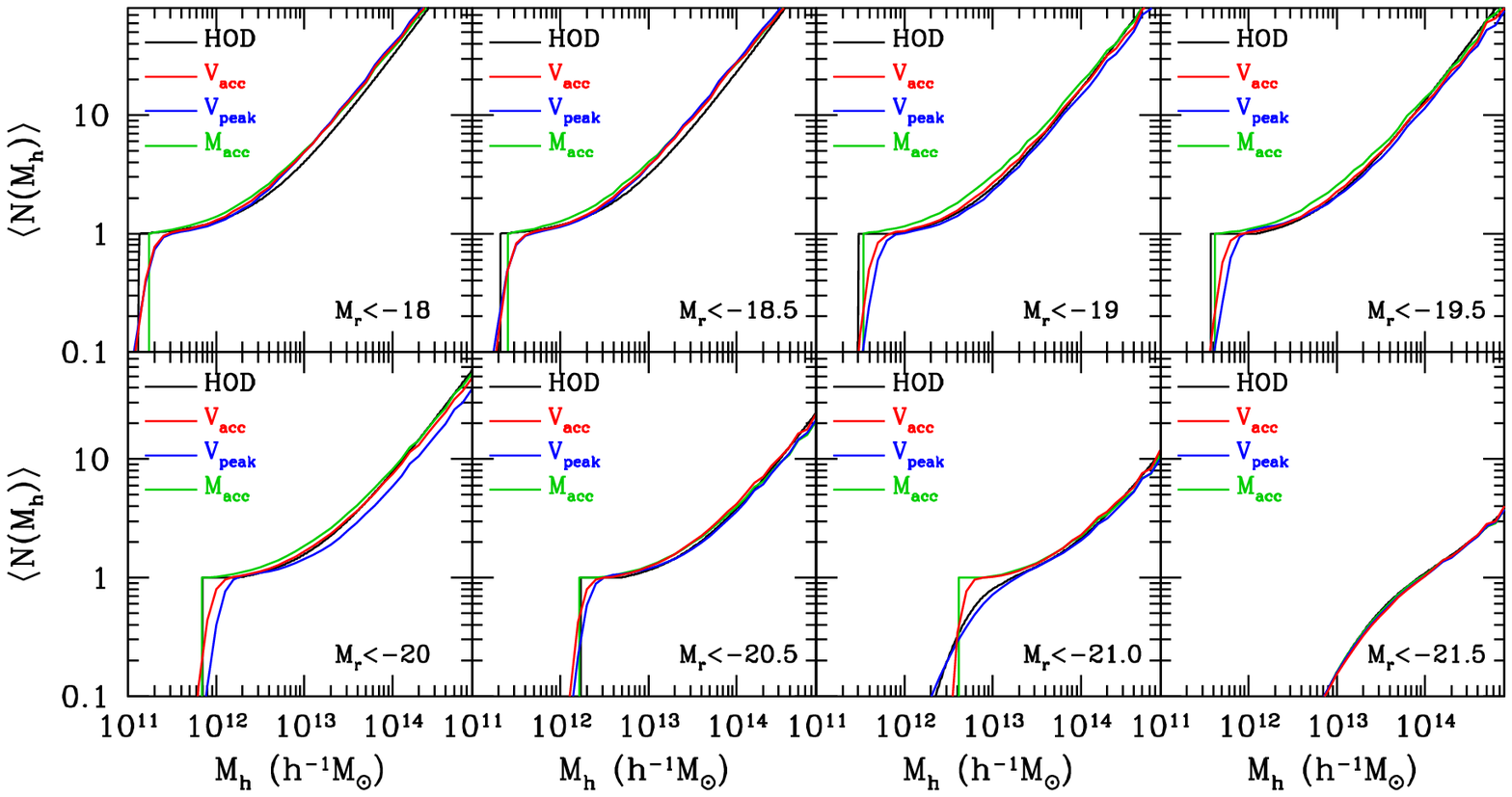}
\caption{Mean halo occupation functions of the best-fitting HOD and SCAM models
from fitting the $w_p$--only data for different luminosity threshold samples.} \label{fig:hodshamc}
\end{figure*}
\begin{figure*}
\includegraphics[width=0.8\textwidth]{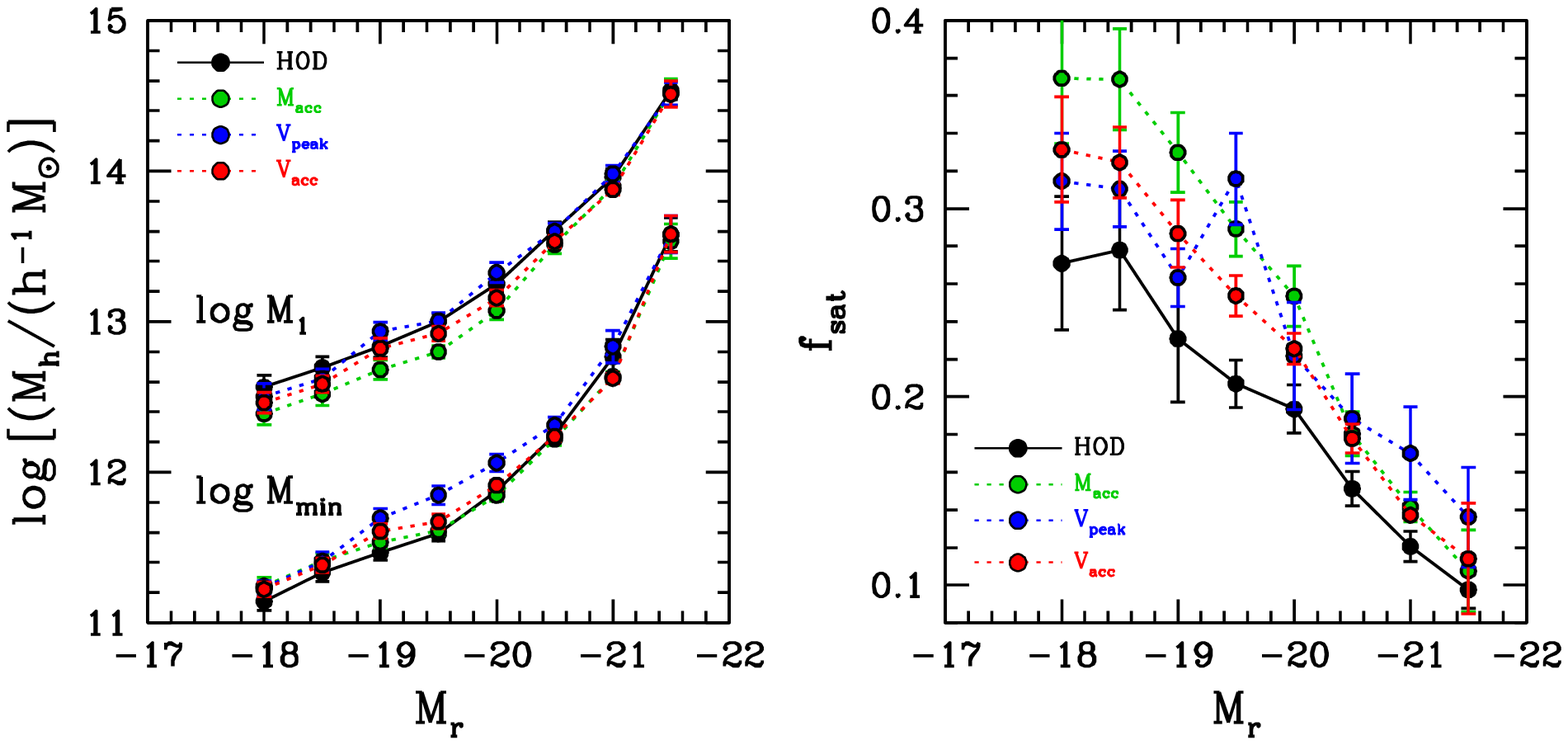}
\caption{Comparisons of the model parameters of the four models from fitting
the $w_p$--only data for the
different luminosity threshold samples. The left-hand panel shows the comparisons
of the characteristic cutoff mass $M_{\rm min}$ of host haloes and the
characteristic mass $M_1$ of
haloes hosting on average one satellite galaxy. The satellite fraction
$f_{\rm sat}$ is shown in the right-hand panel.} \label{fig:m1minwp}
\end{figure*}
However, the goodness of fit to the data cannot be simply judged by eye,
because the full covariance matrices of the measurements need to be taken
into account. Each panel of Fig.~\ref{fig:cov} denotes the normalized
covariance matrix for the corresponding 2PCF measurements shown in
Fig.~\ref{fig:wpsham}. The best-fitting $\chi^2$ for each model is displayed
in Fig.~\ref{fig:chisham}. For example, from Fig.~\ref{fig:wpsham}, it seems
that the $V_{\rm acc}$ model fits slightly better than the $V_{\rm peak}$
model for the $M_r<-19.5$ sample. But the best-fitting $\chi^2$ value of the
$V_{\rm peak}$ model is in fact smaller due to the strong positive
correlation in the neighbouring bins of the data measurements. The large
off-diagonal terms of the covariance matrix are important for all the galaxy
samples except for the most luminous one.

As shown in Fig.~\ref{fig:wpsham} and Fig.~\ref{fig:chisham}, none of the
three SHAM models can provide satisfactory fits for all galaxy samples. The
$V_{\rm peak}$ model fits better for galaxy samples fainter than $-21$, while
the $V_{\rm acc}$ model fits better for more luminous galaxy samples. The
overall goodness-of-fit for the $V_{\rm peak}$ model is around
$\chi^2/\rm{dof}\sim3$. Therefore, the three SHAM models considered above can
hardly be regarded as good models to the observed galaxy projected 2PCFs. We
thus consider the more sophisticated and flexible subhalo models (SCAM) in
the following section.

We show in the left panel of Fig.~\ref{fig:siglogl} the comparisons of the
characteristic cutoff circular velocity and the inferred scatters in galaxy
luminosity in haloes with the cutoff circular velocity in the $V_{\rm acc}$
and $V_{\rm peak}$ models, respectively. The more luminous galaxy samples
have higher cutoff velocities, and the inferred cutoff for $V_{\rm peak}$ is
generally about 0.1 dex higher than that for $V_{\rm acc}$.

As discussed in \S\ref{subsec:sham}, the scatter $\sigma_{\log L}$ in galaxy
luminosity at fixed circular velocity is encoded in the $\sigma_{\log V}$
parameter (width of the cutoff profile in the galaxy occupation function).
Following \citet{Zheng07} (see details in their Eq.~4), we have $\sigma_{\log
L}=p\,\sigma_{\log V}/\sqrt{2}$, where $p$ is the local power-law slope of
the $L$--$V$ relation, i.e. $p\equiv d\log L/d\log V$. To obtain the local
power-law slope, we make use of the formula proposed by \cite{Vale06} to fit
the relation between the sample luminosity threshold $L$ and the velocity
cutoff $V$, ($V_{\rm acc}$ or $V_{\rm peak}$),
\begin{equation}
L=L_0\frac{(V/V_t)^a}{\left[1+(V/V_t)^{bk}\right]^{1/k}}.
\end{equation}
The variables $L_0$, $V_t$, $a$, $b$ and $k$ are the model parameters. As
seen from the left panel of Fig.~\ref{fig:siglogl}, $L$--$V$ can also be well
described by broken power laws, which justifies the use of local power-law
slope $p$ in the above equation. The resulting scatter $\sigma_{\log L}$ is
shown in the right panel of Fig.~\ref{fig:siglogl}. Most scatters are smaller
than 0.3, and the scatters in the $V_{\rm peak}$ model are generally larger.
We note that the uncertainties on the scatters of the faint galaxy samples are 
very large. If the scatters are not taken into account in the SHAM models, only
low-luminosity samples can be reasonably fitted. The scatters become important
for luminous galaxies of $M_r<-20.5$. Overall the scatter we infer is
consistent with that in the Tully-Fisher relation.

\subsection{Results from the SCAM and HOD Models}

The large $\chi^2/\rm{dof}$ values of the SHAM models are mostly caused by
the underestimates of the small-scale clusterings. Since the small-scale
galaxy pairs are dominated by the one-halo term, i.e. intra-halo galaxy
pairs, the above underestimate could be an indication that subhaloes are not
complete in representing satellite galaxies towards the centre of host
haloes. Compared to the stellar components of satellite galaxies, subhaloes
in $N$-body simulations are more easily disrupted, especially in the central
regions of the host haloes where the tidal stripping effect is more
significant. Indeed, the differences in the distribution profiles between
subhaloes and satellite galaxies have been seen from $N$-body and hydrodynamic
simulations of the same initial conditions (e.g. Fig.~7 of
\citealt{Weinberg08} and Fig.~2 of \citealt{Vogelsberger14b}).

However, if we work under the implicit assumption adopted in most SHAM models
that satellites can only reside in subhaloes identified in $N$-body
simulations, there is another way to improve the small-scale clustering fitted
by adding additional components to the SHAM models. If we allow the central
and satellite galaxies to have different occupation distributions in the
distinct haloes and subhaloes as in our SCAM models, the deficiency of
small-scale galaxy pairs can be compensated by more satellite galaxies
populating subhaloes in lower mass host haloes. The galaxy number density can
still be preserved by increasing the cutoff mass (or velocity) scale of the
central galaxies. This seems like an extreme model that possibly artificially
increases the fraction of the satellite galaxies, as we allow the relation between central galaxies and distinct haloes and that between satellites and subhaloes to be completely independent of each other in SCAM, which may not be true in reality. But on the other hand,
there is some evidence that the connections of central and distinct haloes
and those of satellite and subhaloes should be different \citep{Yang09,Yang12,Neistein11,Rodriguez-Puebla12,Wetzel12,Watson13}. Within the SHAM
framework, results from our SCAM model that jointly fits the 2PCFs and the
galaxy number density may serve as a probe to the difference between central
and satellite galaxies.

The best-fitting HOD and SCAM models to the projected 2PCF $w_p$ are shown as
the solid lines in Fig.~\ref{fig:wpshamc}. The $\chi^2$ of the model fittings
are displayed in the left panel of Fig.~\ref{fig:chishamc}. All the three
SCAM models have much better best-fitting $\chi^2$ than the SHAM models, with
only three more free parameters. Judged from the best-fitting $\chi^2$
values, the HOD model and the $V_{\rm acc}$ model are the two best models.
For galaxy samples fainter than $M_r=-20$, the values of $\chi^2/{\rm dof}$
of the two models are both around unity. For more luminous galaxies, the HOD
model has a $\chi^2/{\rm dof}\sim1.8$. Note that in the HOD model, we set a
prior by fixing the high mass end slope $\alpha$ of the satellite mean
occupation function to be unity, for the purpose of reducing the number of
parameters to be the same as in the SCAM models. If we also allow $\alpha$ to
vary, the best-fitting value of $\alpha$ for these luminous galaxies is about
$1.15$ and the $\chi^2/\rm{dof}$ would be significantly reduced to values
around unity for the HOD model, as shown in table~2 of G15.
Compared to $\alpha=1$, the higher-than-unity value of $\alpha$ implies that
luminous satellite galaxies tend to populate even more massive haloes. We
also note that due to the strong correlation in the off-diagonal elements of
covariance matrices, the $\chi^2$ cannot be simply judged from the ratios
between the models and data, as explained in the previous sections. For example,
for the faint galaxy sample of $M_r<-19$, the HOD and $V_{\rm acc}$ model has
almost the same $\chi^2$. However, the model predictions for $w_p$ are quite
different.

Except for the $V_{\rm peak}$ model that has a strong variation of $\chi^2$
with the sample luminosity, all other three models can fit the faint galaxy
samples very well. That is, once we allow the central and satellite galaxies
to have different relations to the host haloes and the subhaloes, the
satellite occupation can be adjusted to reproduce the small-scale
clustering. For the most luminous galaxy sample of $M_r<-21.5$, all the four
models have similar best-fitting $\chi^2$ values. As will be shown in the
following, the ratio between the typical subhalo and the host halo mass is
increasing with the galaxy luminosity \citep[see e.g.][]{Guo14}. According to
Fig.~\ref{fig:profile}, this makes the spatial distribution of subhaloes in
the host haloes approach that of the dark matter, which explains why the SCAM
models produce best-fitting $\chi^2$ values more consistent with the HOD
model.

The right panel of Fig.~\ref{fig:chishamc} shows the best-fitting galaxy
number density for the different models. The $V_{\rm peak}$ model has
slightly lower galaxy number densities for the two samples of $M_r<-19$ and
$M_r<-20$, mainly responsible for the larger $\chi^2$ shown in the left
panel. All other three models reproduce the observed galaxy number densities
remarkably well. We note that different from the SHAM models, in the SCAM
models, the number densities of the models are not required to exactly match
those of the galaxy samples, and the discrepancies in the number densities
contribute to the total $\chi^2$. The models tend to find the balance between
fitting the 2PCFs and fitting the sample number densities. However, the contribution of the number density to the total $\chi^2$ is usually small, since a reasonable model that describes well the 2PCFs also predicts a reasonable sample number density. Even for the case with the largest deviation seen in the right panel of Fig.~\ref{fig:chishamc} (the $V_{\rm peak}$ model for the sample of $M_r<-19$), its contribution to the total $\chi^2$ is only 3.7\%.

Fig.~\ref{fig:hodshamc} shows the mean occupation functions of the
best-fitting HOD and SCAM models. The sharp cutoff profiles are shown for the
faint galaxy samples. But we should note that the scatters between the galaxy
luminosity and the halo properties are not well constrained in all models for
faint galaxies (see also G15). The cutoff profiles in the $V_{\rm acc}$ and
$V_{\rm peak}$ models are softened because of the scatter between the
circular velocity and the halo mass \citep[see also Fig.~5 of][]{Conroy06}.
The trends in the mean occupation function with galaxy luminosity in
different models are similar. For the $M_r<-21.5$ sample, the mean occupation
functions from the four models are closely matched, while the differences
become larger for fainter galaxies.

Fig.~\ref{fig:m1minwp} presents the detailed comparisons of the three HOD
parameters, the characteristic host halo mass $M_{\rm min}$, the
characteristic mass of haloes hosting on average one satellite galaxy $M_1$
and the satellite fraction $f_{\rm sat}$. For the purpose of fair
comparisons, we convert the corresponding model parameters in the SCAM models
to those of the HOD model using Eq.~\ref{eq:satmacc} and the corresponding
version for $V_{\rm acc}$ and $V_{\rm peak}$. Except for the $V_{\rm peak}$
model, all the other three models have consistent constraints to the host
halo mass scale $M_{\rm min}$, because $M_{\rm min}$ is mostly constrained by
the sample number density and the large-scale galaxy bias.

As seen in Fig.~\ref{fig:profile}, the subhalo distribution profile in the
host haloes is generally shallower than that of the dark matter distribution.
The small-scale clustering is sensitive to the satellite occupation
distribution, since it is dominated by the one-halo term, i.e. the galaxy
pairs within the same host halo. In order to compensate the shallower profile
and to match the small-scale clustering measurements of $w_p$, the SCAM
models tend to populate satellite galaxies into lower mass haloes than in the
HOD model. In the SCAM models, this is realized by lowering the mass
(velocity) scale and increasing the scatter for populating subhaloes,
compared to the way of populating distinct haloes. As a consequence, the
characteristic mass $M_1$ (left panel of Fig.~\ref{fig:m1minwp}) inferred
from the SCAM models is generally smaller and the satellite fraction $f_{\rm
sat}$ (right panel of Fig.~\ref{fig:m1minwp}) is higher than that from the
HOD model.
\begin{figure*}
	\includegraphics[width=1.0\textwidth]{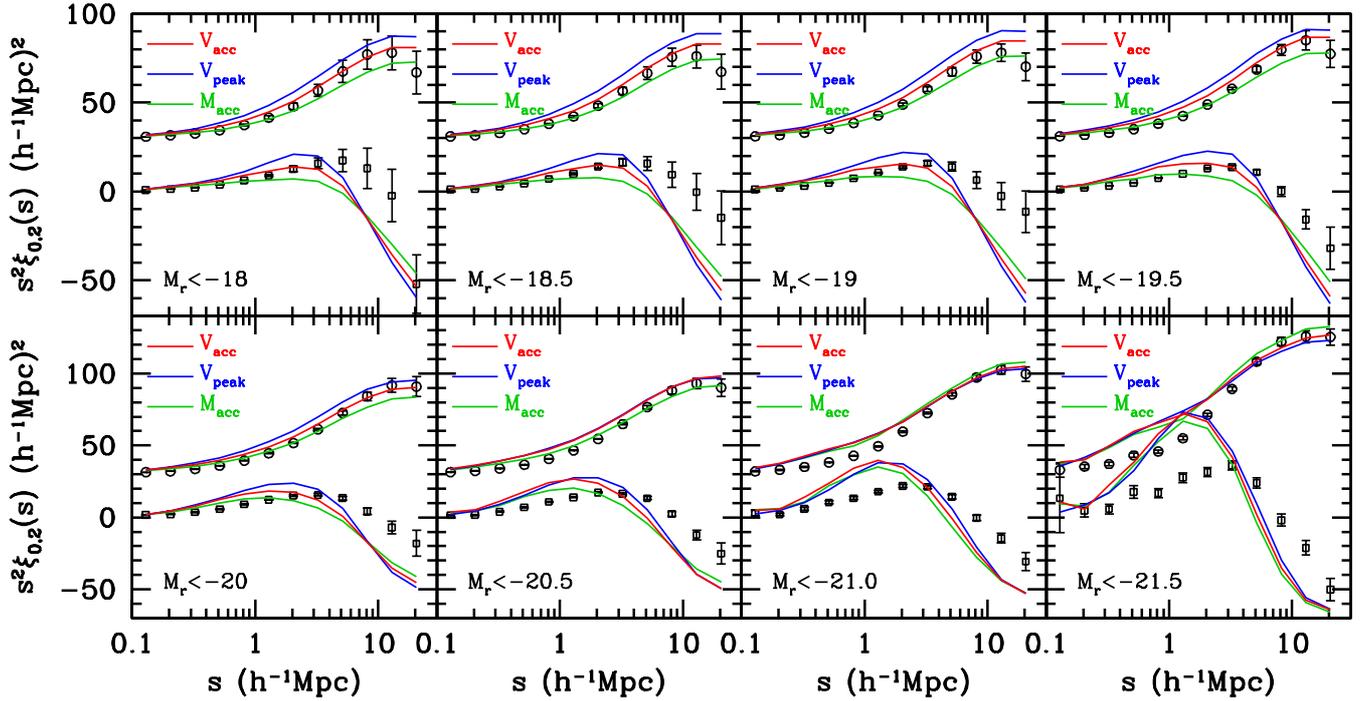}
	\caption{Similar to Fig.~\ref{fig:wpsham}, but for the redshift-space
		monopole (circles) and quadrupole (squares) moments predicted by the SHAM models that best fit $w_p$ only.
		The measured and modelled monopole moments are shifted upwards by 30 for clarity.} \label{fig:xisham}
\end{figure*}
\begin{figure*}
	\includegraphics[width=1.0\textwidth]{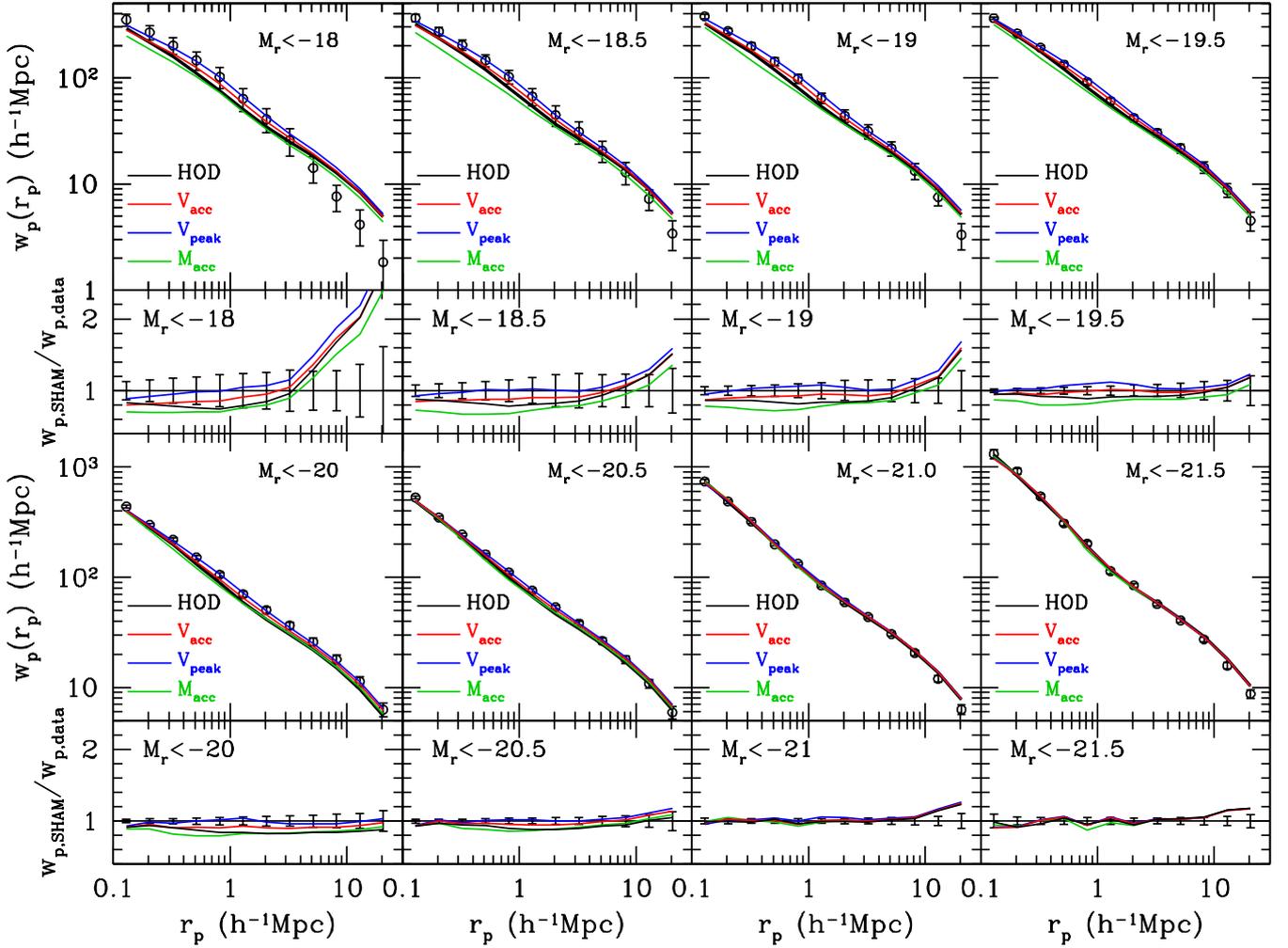}
	\caption{Similar to Fig.~\ref{fig:wpshamc}, but for the bestfitting HOD and SCAM models of fitting both the projected and redshift-space 2PCFs.} \label{fig:wxshamc}
\end{figure*}
\begin{figure*}
\includegraphics[width=1.0\textwidth]{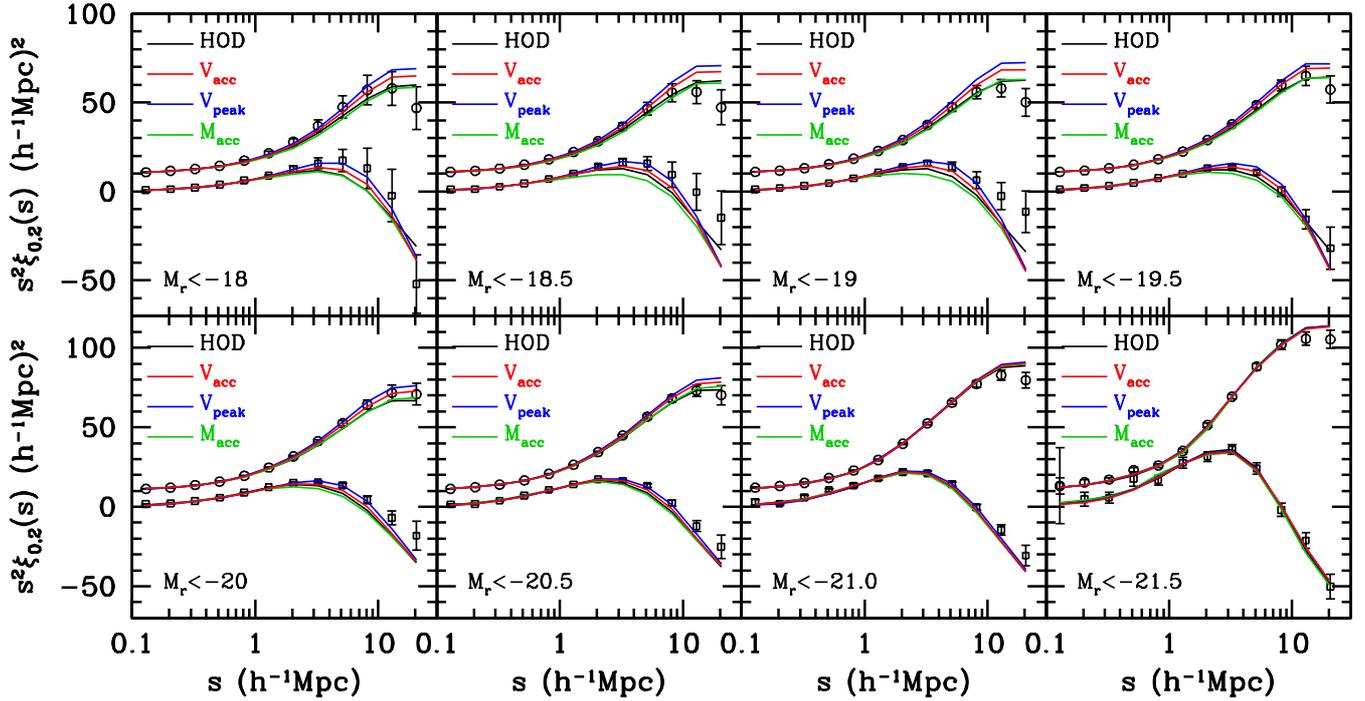}
\caption{Similar to Fig.~\ref{fig:xisham}, but for the HOD and SCAM models. The bestfitting models come from jointly fitting the projected 2PCF $w_p$ and redshift-space 2PCF multiple moments $\xi_{0/2/4}$. The measurements of the
monopole moments are shifted upward by 10 for clarity.} \label{fig:xishamc}
\end{figure*}
\begin{figure*}
\includegraphics[width=0.8\textwidth]{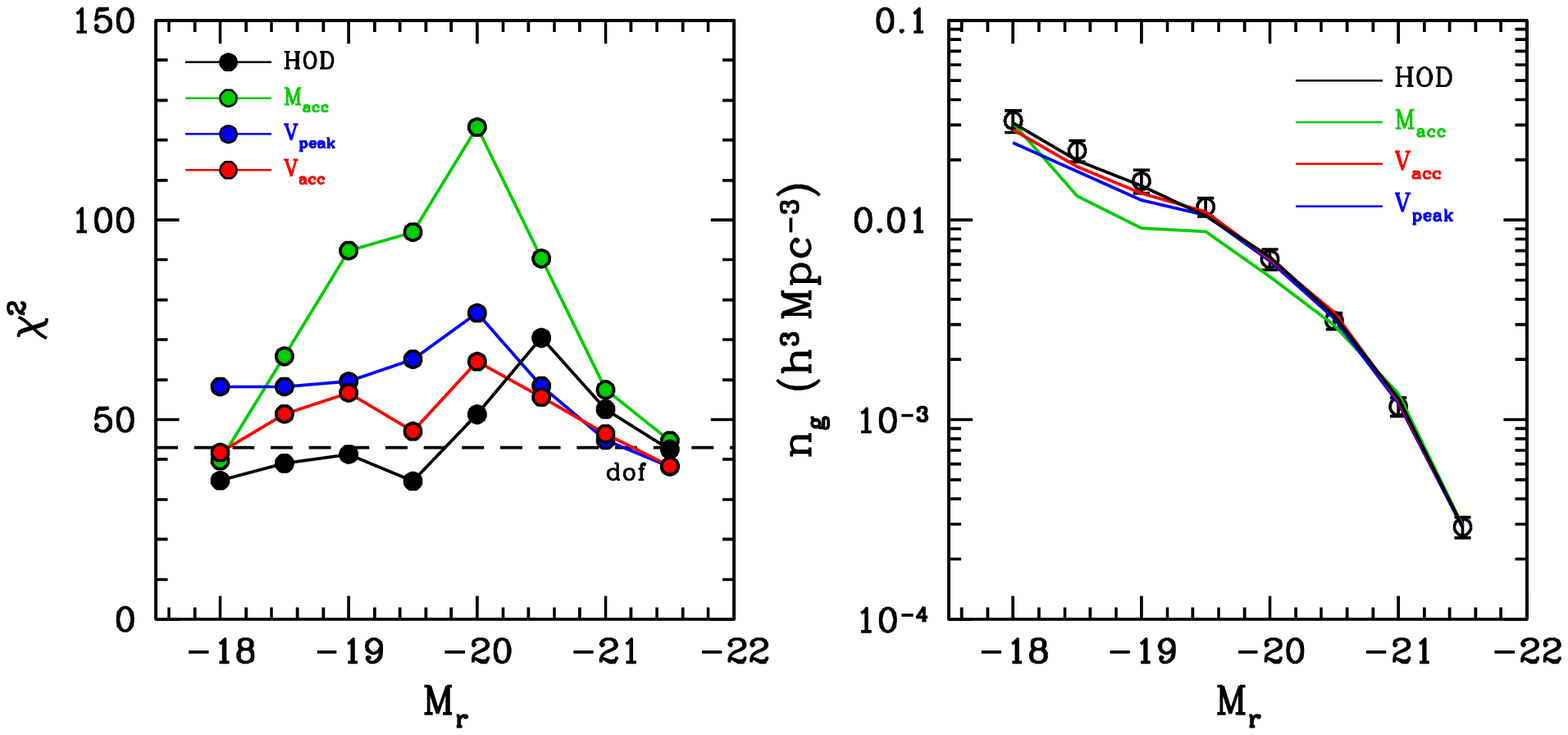}
\caption{Similar to Fig.~\ref{fig:chishamc}, but for models jointly fitting
the projected and redshift-space 2PCFs.} \label{fig:chixsham}
\end{figure*}
The $V_{\rm acc}$ SCAM model shows the best overall agreement with the HOD
model, with more or less consistent best-fitting $\chi^2$ values
(Fig.~\ref{fig:chishamc}). The HOD-related parameters of the four models have
better agreement for luminous galaxies. However, the $\chi^2$ values are
still quite different from model to model (Fig.~\ref{fig:chishamc}),
indicating the effect and importance of the spatial distribution of
satellites (subhaloes or particles in the four models) in modelling
small-scale $w_p$. For example, the model parameters of the three subhalo
models for the $M_r<-20.5$ sample are consistent with each other, but the
$M_{\rm acc}$ model still has a $\chi^2/{\rm dof}$ value as large as 4.2.
Based on the best-fitting $\chi^2$ values, the subhaloes selected by circular
velocities ($V_{\rm acc}$ or $V_{\rm peak}$) seem to better trace the
satellite galaxies \citep[see also e.g.][]{Chaves-Montero15}.

\section{Modelling the Redshift-space 2PCFs}\label{sec:wx}

As shown in G15, jointly fitting the projected and redshift-space 2PCFs helps
tighten the constraints to the galaxy spatial distribution in the haloes, as
well as constraining their velocity distributions. Since the traditional SHAM
models do not have galaxy velocity bias that are required to fit the
redshift-space 2PCFs, the resulting $\chi^2/\rm{dof}$ values are found to be
significantly large. We show in Fig.~\ref{fig:xisham} the predicted redshift-space monopole and quadrupole moments in the SHAM models that bestfit $w_p$. Clearly, the traditional SHAM models fail to describe the redshift-space clustering, especially the quadrupoles. Therefore, in this section, we only compare the HOD and SCAM model fitting results. We first display in Fig.~\ref{fig:wxshamc} the predictions of the projected 2PCF $w_p(r_p)$ for the bestfitting HOD and SCAM models from jointly fitting both the projected and redshift-space 2PCFs. It is similar to Fig.~\ref{fig:wpshamc}, except that the $M_{\rm acc}$ model leads to poorer fits for the faint galaxy samples, as a result of tuning parameters to fit the redshift-space clustering.
Fig.~\ref{fig:xishamc} shows the best fits to the
redshift-space 2PCFs. For clarity, we only show the best-fitting models to
the measured redshift-space monopole (circles) and quadrupole (squares)
moments. The hexadecapole moments are also used in the model fittings, but
not shown in the figure. The $\chi^2$ of the best-fitting models are shown in
the left panel of Fig.~\ref{fig:chixsham}, while the right panel displays the
best-fitting sample number densities.

\begin{figure*}
\includegraphics[width=0.8\textwidth]{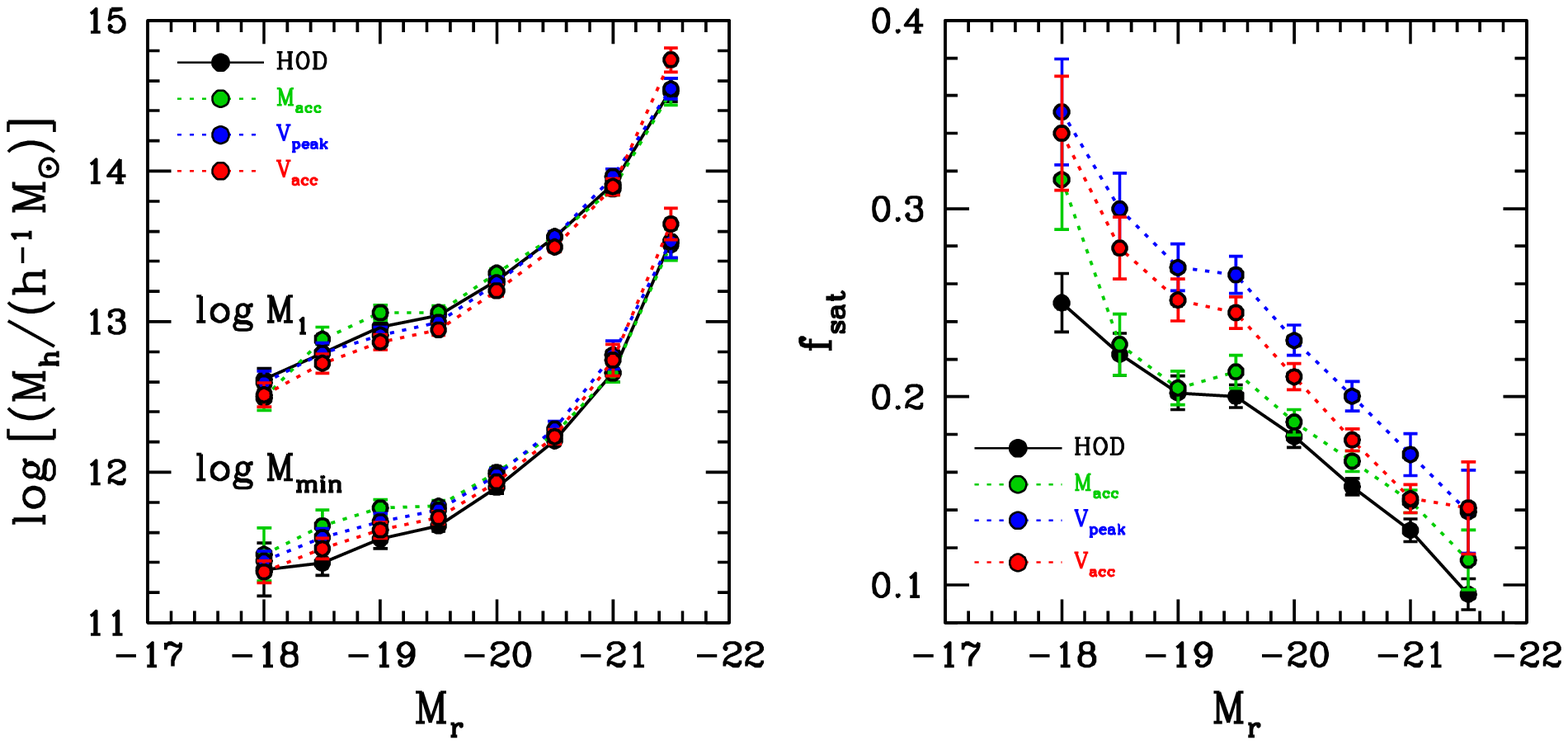}
\caption{Similar to Fig.~\ref{fig:m1minwp}, but for the models jointly
fitting the projected and redshift-space 2PCFs.} \label{fig:m1minwx}
\end{figure*}
\begin{figure*}
\includegraphics[width=1.0\textwidth]{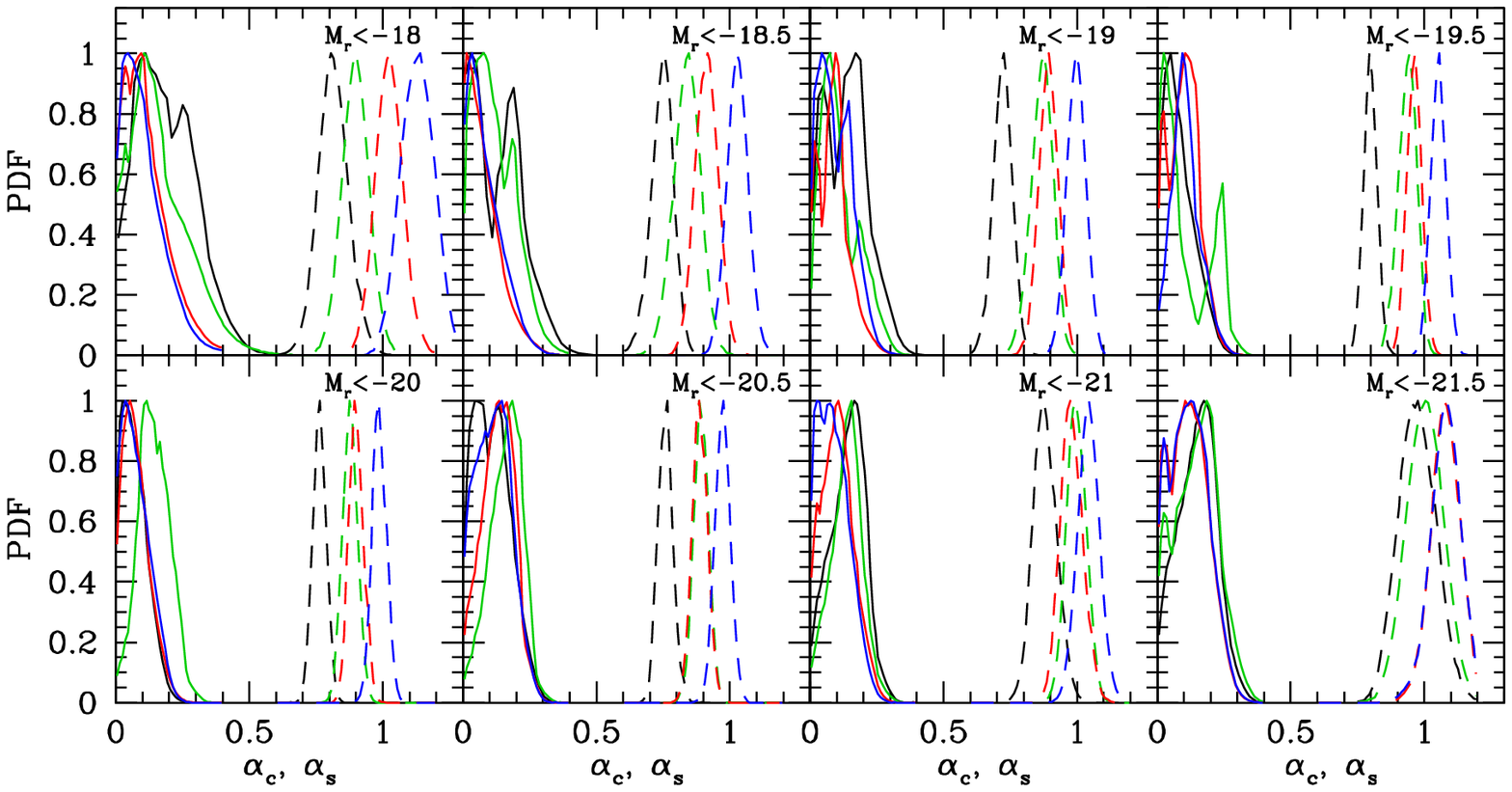}
\caption{Galaxy velocity bias probability distributions for
different models, constrained from jointly fitting the projected and
redshift-space 2PCFs. The solid and dashed lines are for the central
($\alpha_c$)
and satellite ($\alpha_s$) galaxy velocity bias, respectively. Different
panels show the distributions for different luminosity threshold samples. The
black, green, blue and red curves are for the HOD, $M_{\rm acc}$, $V_{\rm
peak}$ and $V_{\rm acc}$ models, respectively.} \label{fig:acsub}
\end{figure*}
Except for the $M_{\rm acc}$ model, all other three models fit the data
reasonably well. As seen from Fig.~\ref{fig:xishamc}, the largest deviation
of the $M_{\rm acc}$ model fits from the measurements and from the fits of
other models lies in the quadrupole, which dominates contributions to the
$\chi^2$. Moreover, the best-fitting sample number densities from the $M_{\rm
acc}$ model are significantly lower than the observed ones for the faint
galaxy samples (except for the $M_r<-18$ sample). Compared to the constraints
from fitting $w_p$ only (Fig.~\ref{fig:chishamc}), the $M_{\rm acc}$ model
has the galaxy number density decreased in the joint-fitting in order to
match the redshift-space clustering. Since the $M_{\rm acc}$ model provides
very good fittings to $w_p$ for the faint galaxies, the failure in matching
the galaxy redshift-space clustering measurements indicates that the
subhaloes selected based on $M_{\rm acc}$ cannot reproduce well the velocity
distribution of the satellite galaxies in the observation.

Except for the sample of $M_r<-20.5$, the HOD model can explain the observed
galaxy 2PCFs very well, with a reasonable $\chi^2/{\rm dof}$ for each sample.
As mentioned in the previous section, the model fitting to the luminous
galaxy samples (including $M_r<-20.5$) can be significantly improved when we
allow the high-mass end slope $\alpha$ of the mean occupation function to
vary \citep[see e.g. Table 2 of][]{Guo15c}. Among the three SCAM models, the
$V_{\rm acc}$ model better fits the data than the other two subhalo models,
similar to the case of fitting $w_p$ only. The dof of the models is 43 (48
2PCF data points plus one number density and minus six free parameters), and
the 2$\sigma$ range of the expected $\chi^2$ distribution is about $43\pm
18.5$. Even though the $\chi^2$ values from the HOD model are overall lower
than those from the subhalo models, those from the $V_{\rm acc}$ and $V_{\rm
peak}$ models are still within the 2$\sigma$ range, giving reasonable fits to
the data.

Fig.~\ref{fig:m1minwx} shows comparisons of the parameters of $M_1$, $M_{\rm
min}$ and $f_{\rm sat}$, as in Fig.~\ref{fig:m1minwp}. Similar to the results
from fitting $w_p$ only, differences in $M_{\rm min}$ and $M_1$ from
different models become larger for fainter galaxy samples. If we focus on
comparing the HOD model and the $V_{\rm acc}$ and $V_{\rm peak}$ subhalo
models (that provide reasonable fits to the data), we find that the HOD
model has the smallest $M_{\rm min}$ and highest $M_1$ values, and the lowest
satellite fraction. The SCAM models tend to populate satellite galaxies into
lower mass haloes to compensate their shallower spatial distribution in the
host haloes. Compared to the right panel of Fig.~\ref{fig:m1minwp}, the
uncertainties in $f_{\rm sat}$ are greatly reduced, because the
redshift-space clustering puts more constraints on the satellite galaxy
distributions.

\begin{figure}
\includegraphics[width=0.42\textwidth]{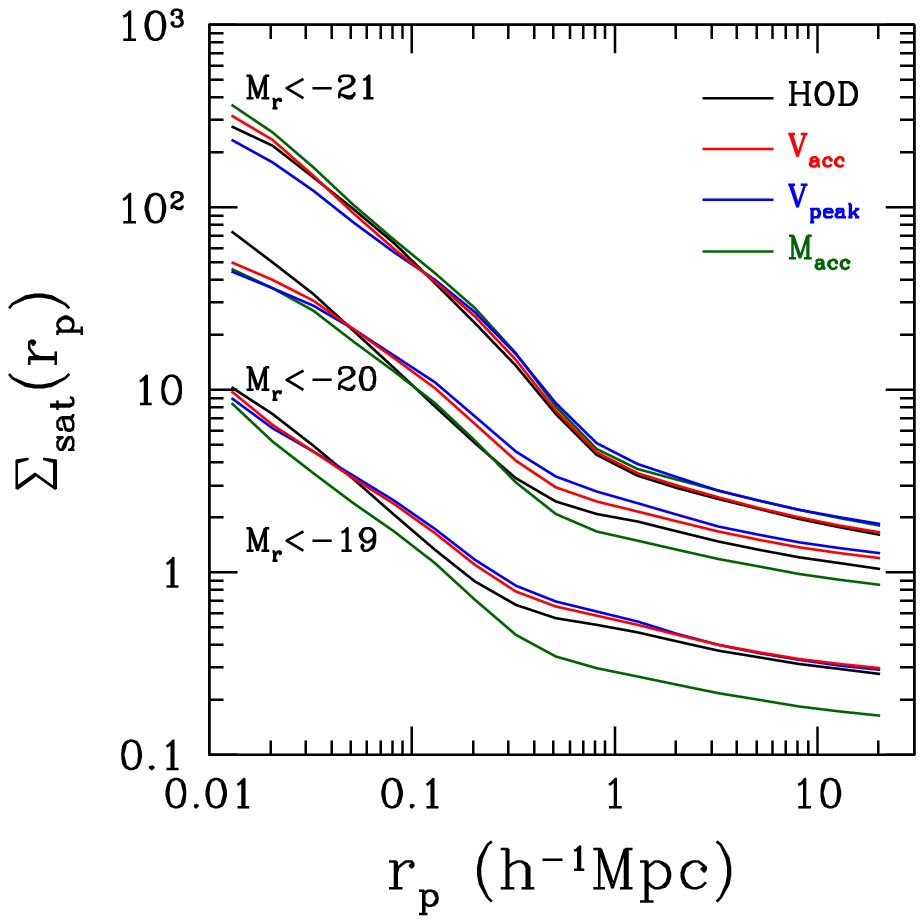}
\caption{Projected number density profile for satellite galaxies from the four different bestfitting models. Offsets are added to separate the cases of different luminosity threshold samples for clarity.} \label{fig:prosham}
\end{figure}
We show in Fig.~\ref{fig:acsub} the model constraints to the galaxy velocity
bias parameters for the different luminosity threshold samples. The black,
green, blue and red curves are for the HOD, $M_{\rm acc}$, $V_{\rm peak}$ and
$V_{\rm acc}$ models, respectively. The solid and dashed lines are for the
central ($\alpha_c$) and satellite ($\alpha_s$) galaxy velocity bias
parameters, respectively. The model constraints for the central galaxy
velocity bias are generally consistent with each other. The best-fitting
$\alpha_c$ values are much smaller than those in G15. The difference is
caused by the different reference to define the velocity bias. In this paper,
the reference halo velocity is defined as the average particle velocities
within inner 10\% halo radius (core), while the velocity bias $\alpha_c$ in
G15 is with respect to the halo bulk velocity. There is a relative motion
between the core and bulk of a halo \citep{Behroozi13}. An average central
galaxy velocity bias $\alpha_c\sim0.1$ is required to fit the redshift-space
2PCFs.

For the satellite velocity bias $\alpha_s$, the results from the HOD and the
SCAM models cannot be directly compared. The satellite velocity bias
$\alpha_s$ for the HOD model is defined with respect to the dark matter
velocity dispersions within the haloes, i.e. $\alpha_{s,\rm{HOD}}=\sigma_{\rm
sat}/\sigma_v$, while the satellite velocity bias in the SCAM models is with
respect to the velocity dispersions of the subhaloes in the host haloes, i.e.
$\alpha_{s,\rm{SCAM}}=\sigma_{\rm sat}/\sigma_{\rm sub}=(\sigma_{\rm
sat}/\sigma_v)/(\sigma_{\rm sub}/\sigma_v)=\alpha_{s,\rm{HOD}}/\alpha_{\rm
sub}$. The subhalo velocity bias $\alpha_{\rm sub}$ is measured to vary from
$1.02$ to $1.11$ in \S\ref{sec:sub}. We take a medium value of $1.07$ for
$\alpha_{\rm sub}$. So we can directly compare $\alpha_{s,\rm{HOD}}$ and
$\alpha_{\rm sub}\alpha_{s,\rm{SCAM}}$. The value of $\alpha_{s,\rm{HOD}}$ is
around $0.8$ for faint galaxies, and increases with luminosity for the two
most luminous galaxy samples, consistent with the results of G15. But
$\alpha_{s,\rm{HOD}}$ is always smaller than $\alpha_{s,\rm{SCAM}}$ (hence
even smaller than $\alpha_{\rm sub}\alpha_{s,\rm{SCAM}}$) inferred from the
three SCAM models. There are also significant differences in
$\alpha_{s,\rm{SCAM}}$ among the three SCAM models, with the $M_{\rm acc}$
model having the smallest $\alpha_{s,\rm{SCAM}}$ and the $V_{\rm peak}$ model
having the largest. The results manifest that models with a shallower
satellite spatial distribution need a compensation of having more satellites
in lower mass haloes and a larger boost in velocity dispersion to match the
redshift-space distortion, consistent with the test shown in Fig. 11 of
\cite{Guo15a}.

\begin{figure*}
\includegraphics[width=0.8\textwidth]{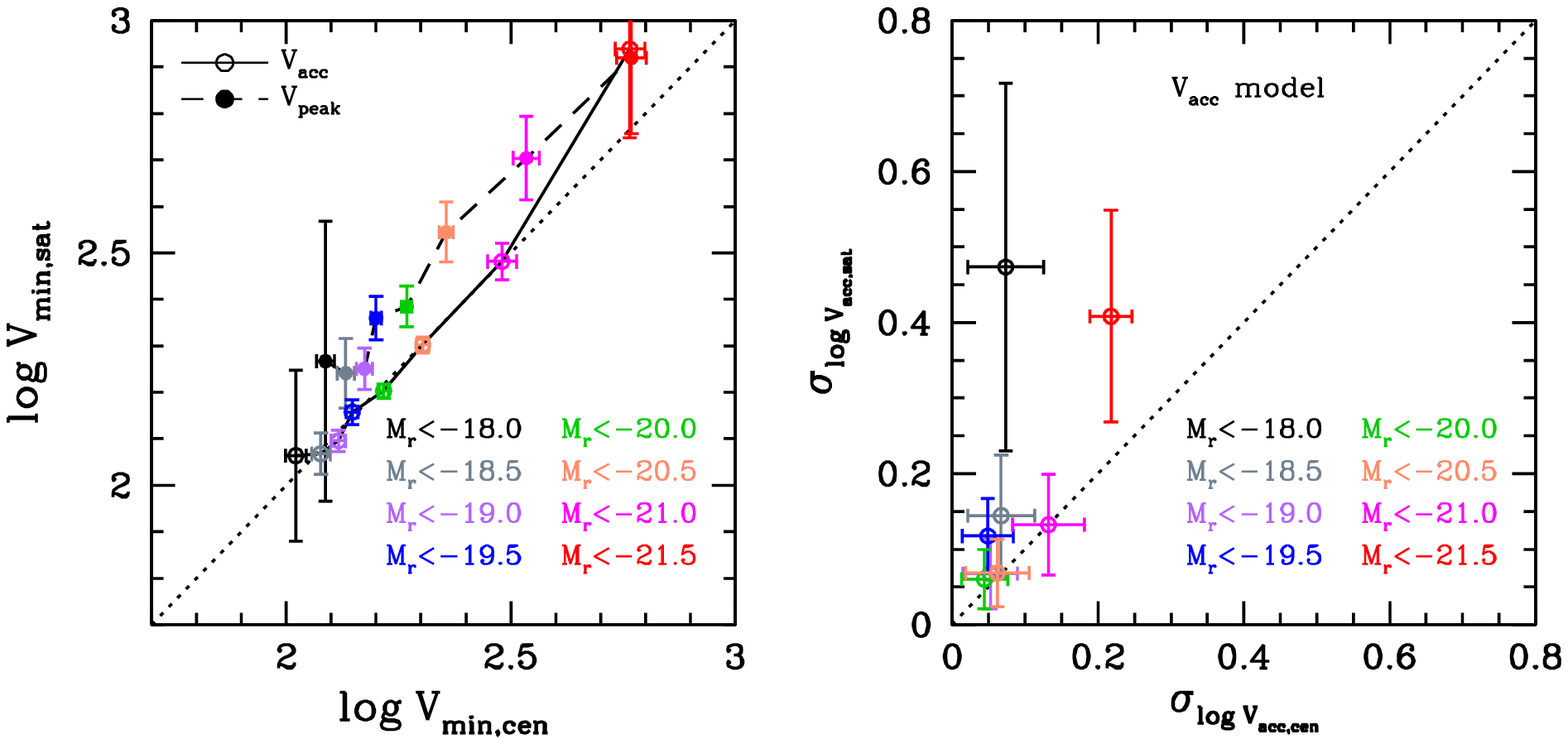}
\caption{Comparisons of the subhalo model parameters for the central and
satellite galaxies from jointly fitting the projected and redshift-space
2PCFs. The left panel shows the comparisons of the circular
velocity thresholds $V_{\rm min,cen}$ and $V_{\rm min,sat}$ for the $V_{\rm
acc}$ (open circles with solid line) and $V_{\rm peak}$ (filled circles with
dashed line) models. The right panel shows the comparisons of the scatters
$\sigma_{\log V_{\rm cen}}$ and $\sigma_{\log V_{\rm sat}}$ for the $V_{\rm
acc}$ model only. See text for details.} \label{fig:mincswx}
\end{figure*}
Satellites in the HOD model have the steepest spatial distribution profile.
Subhaloes in the $M_{\rm acc}$ model have steeper density profile than those
in the other two subhalo models. We show in Fig.~\ref{fig:prosham} three
examples for the projected satellite galaxy number density profiles $\Sigma_{\rm sat}(r_p)$ as a function of the projected distance $r_p$ to centres of hosting haloes \citep[see e.g.][]{Chen06,Wang14}. The projected number density is integrated over the same line-of-sight distance as in the calculation of $w_p(r_p)$, i.e. $40\mpchi$. The turnover points in each sample roughly show the scale of the virial radii of the hosting haloes in these samples.
The trend of the satellite density profiles is
consistent with the behaviour of satellite velocity bias $\alpha_s$ in
Fig.~\ref{fig:acsub}. Although the $M_{\rm acc}$ model generally has a slope of the
satellite galaxy density profile closer to the dark matter distribution, it
does not necessarily lead to better fits to the galaxy 2PCF measurements. The
difference in the different subhalo models is not only in the resulting
subhalo density profiles, but also in the different hosting halo masses (left
panel of Fig.~\ref{fig:m1minwx}). The difference in the satellite density
profiles is partly compensated by the different satellite fraction 
$f_{\rm sat}$ in each model. The $V_{\rm peak}$ model has the highest $f_{\rm sat}$ in each galaxy
sample (right panel of Fig.~\ref{fig:m1minwx}) to compensate for its
shallowest satellite distribution profiles.

Since in our subhalo models we allow the central and satellite galaxies to
have different relations with the hosting haloes (subhaloes), we can compare
the model parameters for the central and satellite galaxies. Since the
$M_{\rm acc}$ model does not have a good best-fitting $\chi^2$ for each
galaxy sample, we focus on the comparisons between the $V_{\rm acc}$ and
$V_{\rm peak}$ models. The left panel of Fig.~\ref{fig:mincswx} shows the
comparisons of the circular velocity thresholds $V_{\rm min, cen}$ and
$V_{\rm min, sat}$ for the $V_{\rm acc}$ (open circles with solid line) and
$V_{\rm peak}$ (filled circles with dashed line) models. It is clear that the
assumption of the same galaxy--halo relation for central and satellite
galaxies does not hold for the $V_{\rm peak}$ model, where $V_{\rm min,sat}$
is generally much larger than $V_{\rm min,cen}$. However, the $V_{\rm acc}$
model has almost the same circular velocities for central and satellite
galaxies. The relation that $V_{\rm min,cen}=V_{\rm min,sat}$ holds within
errors for all the luminosity threshold samples.

The right panel of Fig.~\ref{fig:mincswx} shows the scatter parameter
$\sigma_{\log V_{\rm acc}}$ in the $V_{\rm acc}$ model for distinct haloes
and subhaloes (corresponding to central and satellite galaxies). In general,
the scatters for the central and satellite galaxies are not equal to each
other, with the satellite galaxies having larger scatters between the
luminosity and $V_{\rm acc}$. For the three luminosity threshold samples
around $L_*$, i.e. $M_r<-20$, $-20.5$ and $-21$, central and satellite
galaxies have similar $V_{\rm min, acc}$ and $\sigma_{\log V_{\rm acc}}$. It
implies that the SHAM model with scatter works well for these samples, which
is consistent with the low $\chi^2$ values in the $V_{\rm acc}$ model of
$w_p$-only data in Fig.~\ref{fig:chisham}. But for other samples, central and
satellite galaxies have different scatters in the luminosity-velocity
relation, with satellites having larger scatters, which may be interpreted as
resulted from the different evolution histories of the central and satellite
galaxies.

We note that the $V_{\rm peak}$ model generally has a higher $V_{\rm
min,sat}$ than $V_{\rm min,cen}$, compared to the $V_{\rm acc}$ model.
However, the $V_{\rm peak}$ model has a higher satellite fraction $f_{\rm
sat}$ (right panel of Fig.~\ref{fig:m1minwx}), owing to a much larger
satellite luminosity-velocity scatter ($\sigma_{\log V_{\rm peak,sat}}$) than
in the $V_{\rm acc}$ model.

As a whole, when modelling redshift-space 2PCFs, we find that both the HOD and
SCAM models can give reasonable fits to the measurements for luminous galaxy
samples (above $L_*$). For low luminosity galaxy samples (below $L_*$), the
HOD model, which use dark matter particles to represent satellite galaxies,
leads to the lowest $\chi^2$ among all the models. Among the subhalo models,
if the best-fitting $\chi^2$ values of low luminosity samples are compared,
the $V_{\rm acc}$ model has the best performance. The $V_{\rm peak}$ model is
somewhat worse, and the $M_{\rm acc}$ model just fails to fit the data
(except for the $M_r<-18$ sample). The results imply that the circular
velocities $V_{\rm acc}$ and $V_{\rm peak}$ are more correlated with
satellite luminosity than $M_{\rm acc}$.

\section{Conclusions and Discussions}\label{sec:conclusion}
In this paper, we employ the HOD model and different SHAM models (and the
extension, the SCAM models) to model the projected and redshift-space 2PCF
measurements for the different luminosity threshold samples in the SDSS DR7
Main galaxy sample. All the models are based on the high-resolution
MDPL/SMDPL $N$-body simulations, using the accurate and efficient method
developed in \citet{Zheng16}. We explicitly compare the best-fitting $\chi^2$
values and the modelling results of the HOD model, the SHAM models, and the
SCAM models. The HOD model uses dark matter particles in host haloes to
represent satellite galaxies, while the three sets of SHAM/SCAM models use
halo properties $M_{\rm acc}$, $V_{\rm acc}$, and $V_{\rm peak}$ to establish
the connection between haloes and galaxies, respectively.

In the SHAM model, distinct haloes and subhaloes are treated in the same way
when connected to galaxies. Even with the projected 2PCF $w_p$ data alone, the
SHAM model, no matter which halo property is used, generally fails to provide
satisfactory explanations to all the luminosity threshold samples, with a
typical $\chi^2/{\rm dof}>2$. We therefore introduce the SCAM model by
allowing the relation between central galaxies and distinct haloes and that
between satellite galaxies and subhaloes to be different, and determine the
model parameters by jointly fitting the observed 2PCFs and the sample number
density. The SCAM models give significantly better $\chi^2$ than the SHAM
models.

For an easy comparison, we choose parametrizations so that the HOD and SCAM
models have the same dof. The main difference between the two
models lies in the spatial distribution profile of satellites inside distinct
haloes. Subhaloes (satellite tracers in the SCAM models) generally have a
shallower spatial distribution profile than dark matter particles (satellite
tracers assumed in our HOD model). The shallow distribution profile of
subhaloes in $N$-body simulations may be partially an effect of ignoring the
baryon components --- satellites traced by the more tightly bounded stellar
component are less suffered from tidal disruption that destructs a fraction
of subhaloes near the halo centre. This is supported by the comparisons of
distributions of subhaloes and satellite galaxies in hydrodynamic and
$N$-body simulations \citep[e.g.][]{Weinberg08,Vogelsberger14a}, and
additional investigations along such a direction can shed further light on
such a phenomenon. In this paper, we work under the SHAM assumption that
satellites are traced by subhaloes and investigate to what extent the subhalo
models can interpret the data and to study the corresponding implications.

As expected, the differences in the modelling results between the HOD and SCAM
models and among the different SCAM models can be largely traced back to the
differences in the spatial distribution profile of satellites. Compared to
the HOD modelling results, the SCAM models tend to populate more satellites
into lower mass host haloes to compensate the shallower subhalo distribution
profile and hence to fit the small-scale clustering measurements. This leads
to higher satellite fraction in the SCAM models. When fitting the
redshift-space 2PCFs, we include the central and satellite galaxy velocity
biases in all the models. The derived nonzero central galaxy velocity bias
constraints of the SCAM models are consistent with the HOD model. The
satellite galaxy velocity bias is higher in the SCAM models. The reason is as
follows. As mentioned above, to match the small-scale (real-space)
clustering, more satellites are populated into lower mass haloes in the SCAM
models, and in these host haloes satellite moves more slowly than in the HOD
model. The SCAM models therefore need to boost the velocities of satellites
inside host haloes to fit the redshift-space distortion in the data,
especially the Finger-of-God part.

From jointly modelling the projected and redshift-space 2PCFs, we find that
the HOD model has an overall good performance. For luminous samples (above
$L_*$), all SCAM models provide good fits to the data, and the $V_{\rm peak}$
and $V_{\rm acc}$ models even work better than the HOD model in terms of
$\chi^2$ (Fig.~\ref{fig:chixsham}). However, for galaxy samples with
threshold luminosity below $L_*$, the models become divided. The HOD model is
superb, with the lowest $\chi^2$ values. The $M_{\rm acc}$ model fails to fit
the data (except for the sample with the lowest luminosity threshold,
$M_r<-18$). The $V_{\rm acc}$ and $V_{\rm peak}$ models lead to $\chi^2$
values higher than those from the HOD model, with the $V_{\rm acc}$ model
being better. The $\chi^2$ values from the two models are within the
2$\sigma$ range of the expected value. The results suggest that circular
velocities ($V_{\rm acc}$ and $V_{\rm peak}$) are better quantities than mass
$M_{\rm acc}$ to connect to luminosity of galaxies, especially satellites,
even though $M_{\rm acc}$-selected subhaloes have the steepest spatial
profile among the SCAM models. We therefore recommend that the SHAM model
should no longer use $M_{\rm acc}$ to link to galaxy luminosity. This is in
line with the recent finding by \cite{Contreras15}, who investigate the SHAM
performance for galaxies in two different galaxy formation models and find
that subhalo mass is not a good indicator of galaxy properties. For the two
circular velocity SCAM models, the $V_{\rm acc}$ model is slightly better
than the $V_{\rm peak}$ model in reproducing the projected and redshift-space
2PCFs. In either model, different galaxy--halo relations for central and
satellite galaxies (distinct haloes and subhaloes) are overall required by
the data.

The comparisons between the best-fitting $\chi^2$ for the HOD and SCAM models
show that the HOD model is generally the best model to describe the galaxy
distribution in both projected and redshift spaces. However, the $V_{\rm
acc}$ and $V_{\rm peak}$ models are still acceptable, especially to model
luminous galaxy samples. Including other clustering statistics (e.g. the
three-point correlation functions; \citealt{Guo15b}) may help to further
distinguish these models, as well as to tighten parameter constraints.

It is worth noting that we adopt specific functional forms (Equations \ref{eqn:Ncen} and \ref{eq:macc}) to describe the occupation functions of central and satellite galaxies in the haloes for all the models considered in this paper. Such a functional form is motivated by the results in the semi-analytic models and hydrodynamic simulations of galaxy formation \citep{Zheng05}. It can be derived by assuming a lognormal distribution of the central galaxy luminosity at fixed halo mass and a power-law relation between the mean luminosity of central galaxies and the host halo mass \citep{Zheng07}. In the halo mass range
where the luminosity-halo mass relation (LHMR) or SHMR deviates significantly from a power law, the functional form is less accurate and the interpretation of parameters like $M_{\rm min}$ becomes subtle. \cite{Leauthaud11} compared the difference between the  bestfitting HOD parameter $M_{\rm min}$ (defined as $\langle N_{\rm cen}(M_{\rm min})\rangle=0.5$) with the SHMR of \cite{Behroozi10} and that with a power law SHMR, and found that the difference in $M_{\rm min}$ is $<20\%$ for models with $M_{\rm min}$ in the range of $10^{12}$--$10^{14}M_{\odot}$. For the relevant samples we model, the changes in $\log M_{\rm min}$ are 0.08, 0.04, and -0.04 dex for $M_r<-20.5$, $-21$, and $-21.5$, repectively, all within the 1$\sigma$ model uncertainties.

To derive the functional form of Equation~\ref{eqn:Ncen}, the scatter in 
central galaxy luminosity needs to be independent of halo mass and $\sigma_{\log M_{\rm h}}$ is connected to the luminosity scatter and the form of LHMR. 
In general, $\sigma_{\log M_{\rm h}}$ should not be interpreted as the scatter 
of halo mass at fixed galaxy luminosity \citep{Zheng07,Leauthaud11}. Instead, it describes the width of the cutoff profile of the central galaxy mean 
occupation function, as noted in Section~\ref{subsec:sham}. In modelling the
data, the role of the cutoff profile is to convolve with halo mass function and 
halo bias factor to try to reproduce the galaxy number density and the large-scale galaxy bias, and the two quantities are not sensitive to the functional 
form of the cuttoff profile (as long as the freedoms in width and mass scale 
are included). Therefore, while the interpretation of the parameters like 
$\sigma_{\log M_{\rm h}}$ can be subtle, the modelling results would not be 
affected much by the functional form.

In the implementation of the HOD model, we make the assumption that satellite
galaxies follow the spatial distribution of the dark matter inside haloes.
Although this assumption is commonly adopted in HOD modelling of galaxy
clustering and is loosely motivated by theoretical studies
\citep[e.g.][]{Nagai05}, it needs to be further tested. In hydrodynamic
galaxy formation models, the spatial profile of satellite galaxies depends on
the implementation details. For example, stellar mass loss can be different
for satellites in models with galactic winds of different strengths
\citep[e.g.][]{Simha12}, leading to differences in the spatial distribution
profile of satellites for a given stellar mass threshold (or galaxy number
density). Given such uncertainties, in modelling galaxy clustering, one can
introduce freedom in satellite spatial profile and galaxy formation models
can help inform the sensible parametrization of such a profile.

More generally, comparison of the spatial distributions of satellites, dark
matter, and subhaloes in hydrodynamic and $N$-body simulations can also help
to evaluate the limitations of each model, to improve the prescriptions of
each model, and to choose the best one to model the clustering for a given
sample of galaxies. The validity of the SHAM method can also be tested with
such simulations. \cite{Simha12} applied the SHAM model (with $M_{\rm acc}$
as the halo/subhalo variable) to collisionless $N$-body simulations and
compared with the galaxies in corresponding hydrodynamic simulations (with
the same initial conditions). They find good agreement for the HODs and
satellite distribution profiles for galaxy samples defined by thresholds in
stellar mass. They also find that SHAM slightly overpopulates massive haloes
and hence overpredicts the small-scale clustering, which is attributed to
stellar mass loss of satellite galaxies. The trend seems to be opposite to
our results, although the details depends on the implementation in the
strength of galactic winds. \cite{Chaves-Montero15} also investigate the SHAM
model with $N$-body and the hydrodynamical simulation (the EAGLE simulation)
for stellar mass threshold galaxy samples, using various circular velocities
as the halo/subhalo variables. They found that the peak circular velocity of
a subhalo after relaxation, which is a modified version of the $V_{\rm peak}$
used in our models, correlates most strongly with the galaxy stellar mass.
The SHAM model using this parameter shows better agreement with the galaxy
clustering measurements in the hydrodynamic simulations. Further
investigations following the above ones will be useful (e.g. for
luminosity-threshold samples).

One basic assumption of the HOD model is that the statistical properties of
the galaxy content in a halo only depend on the halo mass. Since the
clustering of haloes of the same mass depends on the halo assembly history
\citep[e.g.][]{Gao05,Wechsler06,Zhu06,Jing07}, the above assumption means
that the halo assembly effect is not translated into galaxy properties in
haloes of the same mass. If the galaxy assembly effect exists (meaning that galaxy properties are correlated with halo assembly), 
it would possibly affect the HOD modelling \citep[e.g.][]{Zu08,Zentner14,Hearin15,Paranjape15} and the
current HOD framework would then need to be extended. However, there is no definite conclusion
yet on whether the assembly bias in galaxy properties shows up in
hydrodynamic simulations \citep[e.g.][]{Berlind03,Chaves-Montero15} or in
galaxy clustering measurements \citep[e.g.][]{Lin16}. According to the
investigation by \cite{Chaves-Montero15} with hydrodynamic simulations,
modelling (with SHAM) based on certain circular velocity variable can capture
about 50\% of the assembly bias effect in galaxy clustering. Since the SCAM
models with circular velocity we introduce in this paper are still less
successful than the HOD model, it remains to be seen whether the galaxy
assembly effect is significant in real data. In any case, further studies on
galaxy assembly are necessary and we reserve such investigations for future
work.

\section*{Acknowledgements}
We thank the anonymous referee for the constructive and detailed comments that help improve the presentation of this paper. We also thank Y.~P.~Jing for helpful comments. 
This work is supported by the 973 Program (No. 2015CB857003). HG acknowledges the
support of NSFC-11543003 and the 100 Talents Program of the Chinese Academy of Sciences. ZZ
was partially supported by NSF grant AST-1208891 and NASA grant NNX14AC89G.
Support for PSB was provided by a Giacconi Fellowship. IZ acknowledges
support, during her sabbatical in Durham, from STFC through grant
ST/L00075X/1, from the European Research Council through ERC Starting Grant
DEGAS-259586 and from a CWRU ACES+ ADVANCE Opportunity Grant. CC, JC, GF, SG,
AK, FP and SRT acknowledge support from the Spanish MICINNs
Consolider-Ingenio 2010 Programme under grant MultiDark CSD2009-00064, MINECO
Centro de Excelencia Severo Ochoa Programme under grant SEV-2012-0249, and
MINECO grant AYA2014-60641-C2-1-P. GY acknowledges financial support from
MINECO (Spain) under research grants AYA2012-31101 and FPA2012-34694.

We gratefully acknowledge the use of the High Performance Computing Resource
in the Core Facility for Advanced Research Computing at Case Western Reserve
University, the use of computing resources at Shanghai Astronomical
Observatory, and the support and resources from the Center for High
Performance Computing at the University of Utah. The MultiDark database was
developed in cooperation with the Spanish MultiDark Consolider Project
CSD2009-00064. The MultiDark-Planck (MDPL) simulation suite has been
performed in the Supermuc supercomputer at LRZ using time granted by PRACE.

Funding for the SDSS and SDSS-II has been provided by the Alfred P. Sloan
Foundation, the Participating Institutions, the National Science Foundation,
the U.S. Department of Energy, the National Aeronautics and Space
Administration, the Japanese Monbukagakusho, the Max Planck Society, and the
Higher Education Funding Council for England. The SDSS Web Site is
http://www.sdss.org/.

The SDSS is managed by the Astrophysical Research Consortium for the
Participating Institutions. The Participating Institutions are the American
Museum of Natural History, Astrophysical Institute Potsdam, University of
Basel, University of Cambridge, Case Western Reserve University, University
of Chicago, Drexel University, Fermilab, the Institute for Advanced Study,
the Japan Participation Group, Johns Hopkins University, the Joint Institute
for Nuclear Astrophysics, the Kavli Institute for Particle Astrophysics and
Cosmology, the Korean Scientist Group, the Chinese Academy of Sciences
(LAMOST), Los Alamos National Laboratory, the Max-Planck-Institute for
Astronomy (MPIA), the Max-Planck-Institute for Astrophysics (MPA), New Mexico
State University, Ohio State University, University of Pittsburgh, University
of Portsmouth, Princeton University, the United States Naval Observatory, and
the University of Washington.

\bibliographystyle{mnras}

\begin{thebibliography}{}
	\makeatletter
	\relax
	\def\mn@urlcharsother{\let\do\@makeother \do\$\do\&\do\#\do\^\do\_\do\%\do\~}
	\def\mn@doi{\begingroup\mn@urlcharsother \@ifnextchar [ {\mn@doi@}
		{\mn@doi@[]}}
	\def\mn@doi@[#1]#2{\def\@tempa{#1}\ifx\@tempa\@empty \href
		{http://dx.doi.org/#2} {doi:#2}\else \href {http://dx.doi.org/#2} {#1}\fi
		\endgroup}
	\def\mn@eprint#1#2{\mn@eprint@#1:#2::\@nil}
	\def\mn@eprint@arXiv#1{\href {http://arxiv.org/abs/#1} {{\tt arXiv:#1}}}
	\def\mn@eprint@dblp#1{\href {http://dblp.uni-trier.de/rec/bibtex/#1.xml}
		{dblp:#1}}
	\def\mn@eprint@#1:#2:#3:#4\@nil{\def\@tempa {#1}\def\@tempb {#2}\def\@tempc
		{#3}\ifx \@tempc \@empty \let \@tempc \@tempb \let \@tempb \@tempa \fi \ifx
		\@tempb \@empty \def\@tempb {arXiv}\fi \@ifundefined
		{mn@eprint@\@tempb}{\@tempb:\@tempc}{\expandafter \expandafter \csname
			mn@eprint@\@tempb\endcsname \expandafter{\@tempc}}}
	
	\bibitem[\protect\citeauthoryear{{Abazajian}, {Adelman-McCarthy}, {Ag{\"u}eros}
		\& et al.}{{Abazajian} et~al.}{2009}]{Abazajian09}
	{Abazajian} K.~N.,  {Adelman-McCarthy} J.~K.,  {Ag{\"u}eros} M.~A.,   et al.
	2009, \mn@doi [\apjs] {10.1088/0067-0049/182/2/543}, \href
	{http://adsabs.harvard.edu/abs/2009ApJS..182..543A} {182, 543}
	
	\bibitem[\protect\citeauthoryear{{Behroozi}, {Conroy}  \&
		{Wechsler}}{{Behroozi} et~al.}{2010}]{Behroozi10}
	{Behroozi} P.~S.,  {Conroy} C.,   {Wechsler} R.~H.,  2010, \mn@doi [\apj]
	{10.1088/0004-637X/717/1/379}, \href
	{http://adsabs.harvard.edu/abs/2010ApJ...717..379B} {717, 379}
	
	\bibitem[\protect\citeauthoryear{{Behroozi}, {Wechsler}  \& {Wu}}{{Behroozi}
		et~al.}{2013}]{Behroozi13}
	{Behroozi} P.~S.,  {Wechsler} R.~H.,   {Wu} H.-Y.,  2013, \mn@doi [\apj]
	{10.1088/0004-637X/762/2/109}, \href
	{http://adsabs.harvard.edu/abs/2013ApJ...762..109B} {762, 109}
	
	\bibitem[\protect\citeauthoryear{{Berlind} \& {Weinberg}}{{Berlind} \&
		{Weinberg}}{2002}]{Berlind02}
	{Berlind} A.~A.,  {Weinberg} D.~H.,  2002, \mn@doi [\apj] {10.1086/341469},
	\href {http://adsabs.harvard.edu/abs/2002ApJ...575..587B} {575, 587}
	
	\bibitem[\protect\citeauthoryear{{Berlind} et~al.,}{{Berlind}
		et~al.}{2003}]{Berlind03}
	{Berlind} A.~A.,  et~al., 2003, \mn@doi [\apj] {10.1086/376517}, \href
	{http://adsabs.harvard.edu/abs/2003ApJ...593....1B} {593, 1}
	
	\bibitem[\protect\citeauthoryear{{Blanton} et~al.,}{{Blanton}
		et~al.}{2005}]{Blanton05b}
	{Blanton} M.~R.,  et~al., 2005, \mn@doi [\aj] {10.1086/429803}, \href
	{http://adsabs.harvard.edu/abs/2005AJ....129.2562B} {129, 2562}
	
	\bibitem[\protect\citeauthoryear{{Bower}, {Benson}, {Malbon}, {Helly}, {Frenk},
		{Baugh}, {Cole}  \& {Lacey}}{{Bower} et~al.}{2006}]{Bower06}
	{Bower} R.~G.,  {Benson} A.~J.,  {Malbon} R.,  {Helly} J.~C.,  {Frenk} C.~S.,
	{Baugh} C.~M.,  {Cole} S.,   {Lacey} C.~G.,  2006, \mn@doi [\mnras]
	{10.1111/j.1365-2966.2006.10519.x}, \href
	{http://adsabs.harvard.edu/abs/2006MNRAS.370..645B} {370, 645}
	
	\bibitem[\protect\citeauthoryear{{Bryan} \& {Norman}}{{Bryan} \&
		{Norman}}{1998}]{Bryan98}
	{Bryan} G.~L.,  {Norman} M.~L.,  1998, \mn@doi [\apj] {10.1086/305262}, \href
	{http://adsabs.harvard.edu/abs/1998ApJ...495...80B} {495, 80}
	
	\bibitem[\protect\citeauthoryear{{Chaves-Montero}, {Angulo}, {Schaye},
		{Schaller}, {Crain}  \& {Furlong}}{{Chaves-Montero}
		et~al.}{2015}]{Chaves-Montero15}
	{Chaves-Montero} J.,  {Angulo} R.~E.,  {Schaye} J.,  {Schaller} M.,  {Crain}
	R.~A.,   {Furlong} M.,  2015, preprint, \href
	{http://adsabs.harvard.edu/abs/2015arXiv150701948C} {} (\mn@eprint {arXiv}
	{1507.01948})
	
	\bibitem[\protect\citeauthoryear{{Chen}, {Kravtsov}, {Prada}, {Sheldon},
		{Klypin}, {Blanton}, {Brinkmann}  \& {Thakar}}{{Chen} et~al.}{2006}]{Chen06}
	{Chen} J.,  {Kravtsov} A.~V.,  {Prada} F.,  {Sheldon} E.~S.,  {Klypin} A.~A.,
	{Blanton} M.~R.,  {Brinkmann} J.,   {Thakar} A.~R.,  2006, \mn@doi [\apj]
	{10.1086/504462}, \href {http://adsabs.harvard.edu/abs/2006ApJ...647...86C}
	{647, 86}
	
	\bibitem[\protect\citeauthoryear{{Conroy}, {Wechsler}  \& {Kravtsov}}{{Conroy}
		et~al.}{2006}]{Conroy06}
	{Conroy} C.,  {Wechsler} R.~H.,   {Kravtsov} A.~V.,  2006, \mn@doi [\apj]
	{10.1086/503602}, \href {http://adsabs.harvard.edu/abs/2006ApJ...647..201C}
	{647, 201}
	
	\bibitem[\protect\citeauthoryear{{Contreras}, {Baugh}, {Norberg}  \&
		{Padilla}}{{Contreras} et~al.}{2015}]{Contreras15}
	{Contreras} S.,  {Baugh} C.~M.,  {Norberg} P.,   {Padilla} N.,  2015, \mn@doi
	[\mnras] {10.1093/mnras/stv1438}, \href
	{http://adsabs.harvard.edu/abs/2015MNRAS.452.1861C} {452, 1861}
	
	\bibitem[\protect\citeauthoryear{{Croton} et~al.,}{{Croton}
		et~al.}{2006}]{Croton06}
	{Croton} D.~J.,  et~al., 2006, \mn@doi [\mnras]
	{10.1111/j.1365-2966.2005.09675.x}, \href
	{http://adsabs.harvard.edu/abs/2006MNRAS.365...11C} {365, 11}
	
	\bibitem[\protect\citeauthoryear{{Gao}, {De Lucia}, {White}  \&
		{Jenkins}}{{Gao} et~al.}{2004}]{Gao04}
	{Gao} L.,  {De Lucia} G.,  {White} S.~D.~M.,   {Jenkins} A.,  2004, \mn@doi
	[\mnras] {10.1111/j.1365-2966.2004.08098.x}, \href
	{http://adsabs.harvard.edu/abs/2004MNRAS.352L...1G} {352, L1}
	
	\bibitem[\protect\citeauthoryear{{Gao}, {Springel}  \& {White}}{{Gao}
		et~al.}{2005}]{Gao05}
	{Gao} L.,  {Springel} V.,   {White} S.~D.~M.,  2005, \mn@doi [\mnras]
	{10.1111/j.1745-3933.2005.00084.x}, \href
	{http://adsabs.harvard.edu/abs/2005MNRAS.363L..66G} {363, L66}
	
	\bibitem[\protect\citeauthoryear{{Guo}, {White}, {Li}  \&
		{Boylan-Kolchin}}{{Guo} et~al.}{2010}]{Guo10}
	{Guo} Q.,  {White} S.,  {Li} C.,   {Boylan-Kolchin} M.,  2010, \mn@doi [\mnras]
	{10.1111/j.1365-2966.2010.16341.x}, \href
	{http://adsabs.harvard.edu/abs/2010MNRAS.404.1111G} {404, 1111}
	
	\bibitem[\protect\citeauthoryear{{Guo} et~al.,}{{Guo} et~al.}{2011}]{Guo11}
	{Guo} Q.,  et~al., 2011, \mnras, \href
	{http://adsabs.harvard.edu/abs/2011MNRAS.413..101G} {413, 101}
	
	\bibitem[\protect\citeauthoryear{{Guo} et~al.,}{{Guo} et~al.}{2013}]{Guo13}
	{Guo} H.,  et~al., 2013, \mn@doi [\apj] {10.1088/0004-637X/767/2/122}, \href
	{http://adsabs.harvard.edu/abs/2013ApJ...767..122G} {767, 122}
	
	\bibitem[\protect\citeauthoryear{{Guo} et~al.,}{{Guo} et~al.}{2014}]{Guo14}
	{Guo} H.,  et~al., 2014, \mn@doi [\mnras] {10.1093/mnras/stu763}, \href
	{http://adsabs.harvard.edu/abs/2014MNRAS.441.2398G} {441, 2398}
	
	\bibitem[\protect\citeauthoryear{{Guo} et~al.,}{{Guo} et~al.}{2015a}]{Guo15a}
	{Guo} H.,  et~al., 2015a, \mn@doi [\mnras] {10.1093/mnras/stu2120}, \href
	{http://adsabs.harvard.edu/abs/2015MNRAS.446..578G} {446, 578}
	
	\bibitem[\protect\citeauthoryear{{Guo} et~al.,}{{Guo} et~al.}{2015b}]{Guo15b}
	{Guo} H.,  et~al., 2015b, \mn@doi [\mnras] {10.1093/mnrasl/slv020}, \href
	{http://adsabs.harvard.edu/abs/2015MNRAS.449L..95G} {449, L95}
	
	\bibitem[\protect\citeauthoryear{{Guo} et~al.,}{{Guo} et~al.}{2015c}]{Guo15c}
	{Guo} H.,  et~al., 2015c, \mn@doi [\mnras] {10.1093/mnras/stv1966}, \href
	{http://adsabs.harvard.edu/abs/2015MNRAS.453.4368G} {453, 4368}
	
	\bibitem[\protect\citeauthoryear{{Hearin}, {Watson}  \& {van den
			Bosch}}{{Hearin} et~al.}{2015}]{Hearin15}
	{Hearin} A.~P.,  {Watson} D.~F.,   {van den Bosch} F.~C.,  2015, \mn@doi
	[\mnras] {10.1093/mnras/stv1358}, \href
	{http://adsabs.harvard.edu/abs/2015MNRAS.452.1958H} {452, 1958}
	
	\bibitem[\protect\citeauthoryear{{Jing}, {Mo}  \& {B{\"o}rner}}{{Jing}
		et~al.}{1998}]{Jing98}
	{Jing} Y.~P.,  {Mo} H.~J.,   {B{\"o}rner} G.,  1998, \mn@doi [\apj]
	{10.1086/305209}, \href {http://adsabs.harvard.edu/abs/1998ApJ...494....1J}
	{494, 1}
	
	\bibitem[\protect\citeauthoryear{{Jing}, {Suto}  \& {Mo}}{{Jing}
		et~al.}{2007}]{Jing07}
	{Jing} Y.~P.,  {Suto} Y.,   {Mo} H.~J.,  2007, \mn@doi [\apj] {10.1086/511130},
	\href {http://adsabs.harvard.edu/abs/2007ApJ...657..664J} {657, 664}
	
	\bibitem[\protect\citeauthoryear{{Klypin}, {Yepes}, {Gottl{\"o}ber}, {Prada}
		\& {He{\ss}}}{{Klypin} et~al.}{2016}]{Klypin16}
	{Klypin} A.,  {Yepes} G.,  {Gottl{\"o}ber} S.,  {Prada} F.,   {He{\ss}} S.,
	2016, \mn@doi [\mnras] {10.1093/mnras/stw248}, \href
	{http://adsabs.harvard.edu/abs/2016MNRAS.457.4340K} {457, 4340}
	
	\bibitem[\protect\citeauthoryear{{Knebe} et~al.,}{{Knebe}
		et~al.}{2013}]{Knebe13}
	{Knebe} A.,  et~al., 2013, \mn@doi [\mnras] {10.1093/mnras/stt1403}, \href
	{http://adsabs.harvard.edu/abs/2013MNRAS.435.1618K} {435, 1618}
	
	\bibitem[\protect\citeauthoryear{{Kravtsov}, {Berlind}, {Wechsler}, {Klypin},
		{Gottl{\"o}ber}, {Allgood}  \& {Primack}}{{Kravtsov}
		et~al.}{2004}]{Kravtsov04}
	{Kravtsov} A.~V.,  {Berlind} A.~A.,  {Wechsler} R.~H.,  {Klypin} A.~A.,
	{Gottl{\"o}ber} S.,  {Allgood} B.,   {Primack} J.~R.,  2004, \mn@doi [\apj]
	{10.1086/420959}, \href {http://adsabs.harvard.edu/abs/2004ApJ...609...35K}
	{609, 35}
	
	\bibitem[\protect\citeauthoryear{{Leauthaud}, {Tinker}, {Behroozi}, {Busha}  \&
		{Wechsler}}{{Leauthaud} et~al.}{2011}]{Leauthaud11}
	{Leauthaud} A.,  {Tinker} J.,  {Behroozi} P.~S.,  {Busha} M.~T.,   {Wechsler}
	R.~H.,  2011, \mn@doi [\apj] {10.1088/0004-637X/738/1/45}, \href
	{http://adsabs.harvard.edu/abs/2011ApJ...738...45L} {738, 45}
	
	\bibitem[\protect\citeauthoryear{{Leauthaud} et~al.,}{{Leauthaud}
		et~al.}{2012}]{Leauthaud12}
	{Leauthaud} A.,  et~al., 2012, \mn@doi [\apj] {10.1088/0004-637X/744/2/159},
	\href {http://adsabs.harvard.edu/abs/2012ApJ...744..159L} {744, 159}
	
	\bibitem[\protect\citeauthoryear{{Lin}, {Mandelbaum}, {Huang}, {Huang},
		{Dalal}, {Diemer}, {Jian}  \& {Kravtsov}}{{Lin} et~al.}{2016}]{Lin16}
	{Lin} Y.-T.,  {Mandelbaum} R.,  {Huang} Y.-H.,  {Huang} H.-J.,  {Dalal} N.,
	{Diemer} B.,  {Jian} H.-Y.,   {Kravtsov} A.,  2016, \mn@doi [\apj]
	{10.3847/0004-637X/819/2/119}, \href
	{http://adsabs.harvard.edu/abs/2016ApJ...819..119L} {819, 119}
	
	\bibitem[\protect\citeauthoryear{{Moster}, {Somerville}, {Maulbetsch}, {van den
			Bosch}, {Macci{\`o}}, {Naab}  \& {Oser}}{{Moster} et~al.}{2010}]{Moster10}
	{Moster} B.~P.,  {Somerville} R.~S.,  {Maulbetsch} C.,  {van den Bosch} F.~C.,
	{Macci{\`o}} A.~V.,  {Naab} T.,   {Oser} L.,  2010, \mn@doi [\apj]
	{10.1088/0004-637X/710/2/903}, \href
	{http://adsabs.harvard.edu/abs/2010ApJ...710..903M} {710, 903}
	
	\bibitem[\protect\citeauthoryear{{Nagai} \& {Kravtsov}}{{Nagai} \&
		{Kravtsov}}{2005}]{Nagai05}
	{Nagai} D.,  {Kravtsov} A.~V.,  2005, \mn@doi [\apj] {10.1086/426016}, \href
	{http://adsabs.harvard.edu/abs/2005ApJ...618..557N} {618, 557}
	
	\bibitem[\protect\citeauthoryear{{Neistein}, {Weinmann}, {Li}  \&
		{Boylan-Kolchin}}{{Neistein} et~al.}{2011a}]{Neistein11a}
	{Neistein} E.,  {Weinmann} S.~M.,  {Li} C.,   {Boylan-Kolchin} M.,  2011a,
	\mn@doi [\mnras] {10.1111/j.1365-2966.2011.18473.x}, \href
	{http://adsabs.harvard.edu/abs/2011MNRAS.414.1405N} {414, 1405}
	
	\bibitem[\protect\citeauthoryear{{Neistein}, {Li}, {Khochfar}, {Weinmann},
		{Shankar}  \& {Boylan-Kolchin}}{{Neistein} et~al.}{2011b}]{Neistein11}
	{Neistein} E.,  {Li} C.,  {Khochfar} S.,  {Weinmann} S.~M.,  {Shankar} F.,
	{Boylan-Kolchin} M.,  2011b, \mn@doi [\mnras]
	{10.1111/j.1365-2966.2011.19145.x}, \href
	{http://adsabs.harvard.edu/abs/2011MNRAS.416.1486N} {416, 1486}
	
	\bibitem[\protect\citeauthoryear{{Nuza} et~al.,}{{Nuza} et~al.}{2013}]{Nuza13}
	{Nuza} S.~E.,  et~al., 2013, \mn@doi [\mnras] {10.1093/mnras/stt513}, \href
	{http://adsabs.harvard.edu/abs/2013MNRAS.432..743N} {432, 743}
	
	\bibitem[\protect\citeauthoryear{{Onions} et~al.,}{{Onions}
		et~al.}{2012}]{Onions12}
	{Onions} J.,  et~al., 2012, \mn@doi [\mnras]
	{10.1111/j.1365-2966.2012.20947.x}, \href
	{http://adsabs.harvard.edu/abs/2012MNRAS.423.1200O} {423, 1200}
	
	\bibitem[\protect\citeauthoryear{{Paranjape}, {Kova{\v c}}, {Hartley}  \&
		{Pahwa}}{{Paranjape} et~al.}{2015}]{Paranjape15}
	{Paranjape} A.,  {Kova{\v c}} K.,  {Hartley} W.~G.,   {Pahwa} I.,  2015,
	\mn@doi [\mnras] {10.1093/mnras/stv2137}, \href
	{http://adsabs.harvard.edu/abs/2015MNRAS.454.3030P} {454, 3030}
	
	\bibitem[\protect\citeauthoryear{{Peacock} \& {Smith}}{{Peacock} \&
		{Smith}}{2000}]{Peacock00}
	{Peacock} J.~A.,  {Smith} R.~E.,  2000, \mn@doi [\mnras]
	{10.1046/j.1365-8711.2000.03779.x}, \href
	{http://adsabs.harvard.edu/abs/2000MNRAS.318.1144P} {318, 1144}
	
	\bibitem[\protect\citeauthoryear{{Planck Collaboration}}{{Planck
			Collaboration}}{2014}]{Planck14}
	{Planck Collaboration} 2014, \mn@doi [\aap] {10.1051/0004-6361/201321591},
	\href {http://adsabs.harvard.edu/abs/2014A%26A...571A..16P} {571, A16}
		
		\bibitem[\protect\citeauthoryear{{Pujol} et~al.,}{{Pujol}
			et~al.}{2014}]{Pujol14}
		{Pujol} A.,  et~al., 2014, \mn@doi [\mnras] {10.1093/mnras/stt2446}, \href
		{http://adsabs.harvard.edu/abs/2014MNRAS.438.3205P} {438, 3205}
		
		\bibitem[\protect\citeauthoryear{{Reddick}, {Wechsler}, {Tinker}  \&
			{Behroozi}}{{Reddick} et~al.}{2013}]{Reddick13}
		{Reddick} R.~M.,  {Wechsler} R.~H.,  {Tinker} J.~L.,   {Behroozi} P.~S.,  2013,
		\mn@doi [\apj] {10.1088/0004-637X/771/1/30}, \href
		{http://adsabs.harvard.edu/abs/2013ApJ...771...30R} {771, 30}
		
		\bibitem[\protect\citeauthoryear{{Rodr{\'{\i}}guez-Puebla}, {Drory}  \&
			{Avila-Reese}}{{Rodr{\'{\i}}guez-Puebla} et~al.}{2012}]{Rodriguez-Puebla12}
		{Rodr{\'{\i}}guez-Puebla} A.,  {Drory} N.,   {Avila-Reese} V.,  2012, \mn@doi
		[\apj] {10.1088/0004-637X/756/1/2}, \href
		{http://adsabs.harvard.edu/abs/2012ApJ...756....2R} {756, 2}
		
		\bibitem[\protect\citeauthoryear{{Rodr{\'{\i}}guez-Puebla}, {Avila-Reese}  \&
			{Drory}}{{Rodr{\'{\i}}guez-Puebla} et~al.}{2013}]{Rodriguez-Puebla13}
		{Rodr{\'{\i}}guez-Puebla} A.,  {Avila-Reese} V.,   {Drory} N.,  2013, \mn@doi
		[\apj] {10.1088/0004-637X/767/1/92}, \href
		{http://adsabs.harvard.edu/abs/2013ApJ...767...92R} {767, 92}
		
		\bibitem[\protect\citeauthoryear{{Rodriguez-Puebla}, {Behroozi}, {Primack},
			{Klypin}, {Lee}  \& {Hellinger}}{{Rodriguez-Puebla}
			et~al.}{2016}]{Rodriguez-Puebla16}
		{Rodriguez-Puebla} A.,  {Behroozi} P.,  {Primack} J.,  {Klypin} A.,  {Lee} C.,
		{Hellinger} D.,  2016, preprint, \href
		{http://adsabs.harvard.edu/abs/2016arXiv160204813R} {} (\mn@eprint {arXiv}
		{1602.04813})
		
		\bibitem[\protect\citeauthoryear{{Sawala} et~al.,}{{Sawala}
			et~al.}{2015}]{Sawala15}
		{Sawala} T.,  et~al., 2015, \mn@doi [\mnras] {10.1093/mnras/stu2753}, \href
		{http://adsabs.harvard.edu/abs/2015MNRAS.448.2941S} {448, 2941}
		
		\bibitem[\protect\citeauthoryear{{Schaye} et~al.,}{{Schaye}
			et~al.}{2015}]{Schaye15}
		{Schaye} J.,  et~al., 2015, \mn@doi [\mnras] {10.1093/mnras/stu2058}, \href
		{http://adsabs.harvard.edu/abs/2015MNRAS.446..521S} {446, 521}
		
		\bibitem[\protect\citeauthoryear{{Simha}, {Weinberg}, {Dav{\'e}}, {Fardal},
			{Katz}  \& {Oppenheimer}}{{Simha} et~al.}{2012}]{Simha12}
		{Simha} V.,  {Weinberg} D.~H.,  {Dav{\'e}} R.,  {Fardal} M.,  {Katz} N.,
		{Oppenheimer} B.~D.,  2012, \mn@doi [\mnras]
		{10.1111/j.1365-2966.2012.21142.x}, \href
		{http://adsabs.harvard.edu/abs/2012MNRAS.423.3458S} {423, 3458}
		
		\bibitem[\protect\citeauthoryear{{Skibba} et~al.,}{{Skibba}
			et~al.}{2015}]{Skibba15}
		{Skibba} R.~A.,  et~al., 2015, \mn@doi [\apj] {10.1088/0004-637X/807/2/152},
		\href {http://adsabs.harvard.edu/abs/2015ApJ...807..152S} {807, 152}
		
		\bibitem[\protect\citeauthoryear{{Somerville}, {Hopkins}, {Cox}, {Robertson}
			\& {Hernquist}}{{Somerville} et~al.}{2008}]{Somerville08}
		{Somerville} R.~S.,  {Hopkins} P.~F.,  {Cox} T.~J.,  {Robertson} B.~E.,
		{Hernquist} L.,  2008, \mn@doi [\mnras] {10.1111/j.1365-2966.2008.13805.x},
		\href {http://adsabs.harvard.edu/abs/2008MNRAS.391..481S} {391, 481}
		
		\bibitem[\protect\citeauthoryear{{Springel} et~al.,}{{Springel}
			et~al.}{2008}]{Springel08}
		{Springel} V.,  et~al., 2008, \mn@doi [\mnras]
		{10.1111/j.1365-2966.2008.14066.x}, \href
		{http://adsabs.harvard.edu/abs/2008MNRAS.391.1685S} {391, 1685}
		
		\bibitem[\protect\citeauthoryear{{Tinker}, {Weinberg}, {Zheng}  \&
			{Zehavi}}{{Tinker} et~al.}{2005}]{Tinker05}
		{Tinker} J.~L.,  {Weinberg} D.~H.,  {Zheng} Z.,   {Zehavi} I.,  2005, \mn@doi
		[\apj] {10.1086/432084}, \href
		{http://adsabs.harvard.edu/abs/2005ApJ...631...41T} {631, 41}
		
		\bibitem[\protect\citeauthoryear{{Vale} \& {Ostriker}}{{Vale} \&
			{Ostriker}}{2006}]{Vale06}
		{Vale} A.,  {Ostriker} J.~P.,  2006, \mn@doi [\mnras]
		{10.1111/j.1365-2966.2006.10605.x}, \href
		{http://adsabs.harvard.edu/abs/2006MNRAS.371.1173V} {371, 1173}
		
		\bibitem[\protect\citeauthoryear{{Vogelsberger} et~al.,}{{Vogelsberger}
			et~al.}{2014a}]{Vogelsberger14a}
		{Vogelsberger} M.,  et~al., 2014a, \mn@doi [\mnras] {10.1093/mnras/stu1536},
		\href {http://adsabs.harvard.edu/abs/2014MNRAS.444.1518V} {444, 1518}
		
		\bibitem[\protect\citeauthoryear{{Vogelsberger} et~al.,}{{Vogelsberger}
			et~al.}{2014b}]{Vogelsberger14b}
		{Vogelsberger} M.,  et~al., 2014b, \mn@doi [\nat] {10.1038/nature13316}, \href
		{http://adsabs.harvard.edu/abs/2014Natur.509..177V} {509, 177}
		
		\bibitem[\protect\citeauthoryear{{Wang}, {Li}, {Kauffmann}  \& {De
				Lucia}}{{Wang} et~al.}{2006}]{Wang06}
		{Wang} L.,  {Li} C.,  {Kauffmann} G.,   {De Lucia} G.,  2006, \mn@doi [\mnras]
		{10.1111/j.1365-2966.2006.10669.x}, \href
		{http://adsabs.harvard.edu/abs/2006MNRAS.371..537W} {371, 537}
		
		\bibitem[\protect\citeauthoryear{{Wang}, {Yang}, {Mo}  \& {van den
				Bosch}}{{Wang} et~al.}{2007}]{Wang07}
		{Wang} Y.,  {Yang} X.,  {Mo} H.~J.,   {van den Bosch} F.~C.,  2007, \mn@doi
		[\apj] {10.1086/519245}, \href
		{http://adsabs.harvard.edu/abs/2007ApJ...664..608W} {664, 608}
		
		\bibitem[\protect\citeauthoryear{{Wang}, {Sales}, {Henriques}  \&
			{White}}{{Wang} et~al.}{2014}]{Wang14}
		{Wang} W.,  {Sales} L.~V.,  {Henriques} B.~M.~B.,   {White} S.~D.~M.,  2014,
		\mn@doi [\mnras] {10.1093/mnras/stu988}, \href
		{http://adsabs.harvard.edu/abs/2014MNRAS.442.1363W} {442, 1363}
		
		\bibitem[\protect\citeauthoryear{{Watson} \& {Conroy}}{{Watson} \&
			{Conroy}}{2013}]{Watson13}
		{Watson} D.~F.,  {Conroy} C.,  2013, \mn@doi [\apj]
		{10.1088/0004-637X/772/2/139}, \href
		{http://adsabs.harvard.edu/abs/2013ApJ...772..139W} {772, 139}
		
		\bibitem[\protect\citeauthoryear{{Wechsler}, {Zentner}, {Bullock}, {Kravtsov}
			\& {Allgood}}{{Wechsler} et~al.}{2006}]{Wechsler06}
		{Wechsler} R.~H.,  {Zentner} A.~R.,  {Bullock} J.~S.,  {Kravtsov} A.~V.,
		{Allgood} B.,  2006, \mn@doi [\apj] {10.1086/507120}, \href
		{http://adsabs.harvard.edu/abs/2006ApJ...652...71W} {652, 71}
		
		\bibitem[\protect\citeauthoryear{{Weinberg}, {Colombi}, {Dav{\'e}}  \&
			{Katz}}{{Weinberg} et~al.}{2008}]{Weinberg08}
		{Weinberg} D.~H.,  {Colombi} S.,  {Dav{\'e}} R.,   {Katz} N.,  2008, \mn@doi
		[\apj] {10.1086/524646}, \href
		{http://adsabs.harvard.edu/abs/2008ApJ...678....6W} {678, 6}
		
		\bibitem[\protect\citeauthoryear{{Wetzel}, {Tinker}  \& {Conroy}}{{Wetzel}
			et~al.}{2012}]{Wetzel12}
		{Wetzel} A.~R.,  {Tinker} J.~L.,   {Conroy} C.,  2012, \mn@doi [\mnras]
		{10.1111/j.1365-2966.2012.21188.x}, \href
		{http://adsabs.harvard.edu/abs/2012MNRAS.424..232W} {424, 232}
		
		\bibitem[\protect\citeauthoryear{{White} \& {Rees}}{{White} \&
			{Rees}}{1978}]{White78}
		{White} S.~D.~M.,  {Rees} M.~J.,  1978, \mnras, \href
		{http://adsabs.harvard.edu/abs/1978MNRAS.183..341W} {183, 341}
		
		\bibitem[\protect\citeauthoryear{{Wu}, {Hahn}, {Evrard}, {Wechsler}  \&
			{Dolag}}{{Wu} et~al.}{2013}]{Wu13b}
		{Wu} H.-Y.,  {Hahn} O.,  {Evrard} A.~E.,  {Wechsler} R.~H.,   {Dolag} K.,
		2013, \mn@doi [\mnras] {10.1093/mnras/stt1582}, \href
		{http://adsabs.harvard.edu/abs/2013MNRAS.436..460W} {436, 460}
		
		\bibitem[\protect\citeauthoryear{{Yamamoto}, {Masaki}  \& {Hikage}}{{Yamamoto}
			et~al.}{2015}]{Yamamoto15}
		{Yamamoto} M.,  {Masaki} S.,   {Hikage} C.,  2015, preprint, \href
		{http://adsabs.harvard.edu/abs/2015arXiv150303973Y} {} (\mn@eprint {arXiv}
		{1503.03973})
		
		\bibitem[\protect\citeauthoryear{{Yang}, {Mo}  \& {van den Bosch}}{{Yang}
			et~al.}{2003}]{Yang03}
		{Yang} X.,  {Mo} H.~J.,   {van den Bosch} F.~C.,  2003, \mn@doi [\mnras]
		{10.1046/j.1365-8711.2003.06254.x}, \href
		{http://adsabs.harvard.edu/abs/2003MNRAS.339.1057Y} {339, 1057}
		
		\bibitem[\protect\citeauthoryear{{Yang}, {Mo}, {Jing}, {van den Bosch}  \&
			{Chu}}{{Yang} et~al.}{2004}]{Yang04}
		{Yang} X.,  {Mo} H.~J.,  {Jing} Y.~P.,  {van den Bosch} F.~C.,   {Chu} Y.,
		2004, \mn@doi [\mnras] {10.1111/j.1365-2966.2004.07744.x}, \href
		{http://adsabs.harvard.edu/abs/2004MNRAS.350.1153Y} {350, 1153}
		
		\bibitem[\protect\citeauthoryear{{Yang}, {Mo}, {van den Bosch}  \&
			{Jing}}{{Yang} et~al.}{2005}]{Yang05c}
		{Yang} X.,  {Mo} H.~J.,  {van den Bosch} F.~C.,   {Jing} Y.~P.,  2005, \mnras,
		\href {http://adsabs.harvard.edu/abs/2005MNRAS.356.1293Y} {356, 1293}
		
		\bibitem[\protect\citeauthoryear{{Yang}, {Mo}  \& {van den Bosch}}{{Yang}
			et~al.}{2009}]{Yang09}
		{Yang} X.,  {Mo} H.~J.,   {van den Bosch} F.~C.,  2009, \mn@doi [\apj]
		{10.1088/0004-637X/693/1/830}, \href
		{http://adsabs.harvard.edu/abs/2009ApJ...693..830Y} {693, 830}
		
		\bibitem[\protect\citeauthoryear{{Yang}, {Mo}, {van den Bosch}, {Zhang}  \&
			{Han}}{{Yang} et~al.}{2012}]{Yang12}
		{Yang} X.,  {Mo} H.~J.,  {van den Bosch} F.~C.,  {Zhang} Y.,   {Han} J.,  2012,
		\mn@doi [\apj] {10.1088/0004-637X/752/1/41}, \href
		{http://adsabs.harvard.edu/abs/2012ApJ...752...41Y} {752, 41}
		
		\bibitem[\protect\citeauthoryear{{Zehavi} et~al.,}{{Zehavi}
			et~al.}{2011}]{Zehavi11}
		{Zehavi} I.,  et~al., 2011, \mn@doi [\apj] {10.1088/0004-637X/736/1/59}, \href
		{http://adsabs.harvard.edu/abs/2011ApJ...736...59Z} {736, 59}
		
		\bibitem[\protect\citeauthoryear{{Zentner}, {Hearin}  \& {van den
				Bosch}}{{Zentner} et~al.}{2014}]{Zentner14}
		{Zentner} A.~R.,  {Hearin} A.~P.,   {van den Bosch} F.~C.,  2014, \mn@doi
		[\mnras] {10.1093/mnras/stu1383}, \href
		{http://adsabs.harvard.edu/abs/2014MNRAS.443.3044Z} {443, 3044}
		
		\bibitem[\protect\citeauthoryear{{Zheng}}{{Zheng}}{2004}]{Zheng04}
		{Zheng} Z.,  2004, \mn@doi [\apj] {10.1086/421542}, \href
		{http://adsabs.harvard.edu/abs/2004ApJ...610...61Z} {610, 61}
		
		\bibitem[\protect\citeauthoryear{{Zheng} \& {Guo}}{{Zheng} \&
			{Guo}}{2016}]{Zheng16}
		{Zheng} Z.,  {Guo} H.,  2016, \mn@doi [\mnras] {10.1093/mnras/stw523}, \href
		{http://adsabs.harvard.edu/abs/2016MNRAS.458..4015Z} {458, 4015}
		
		\bibitem[\protect\citeauthoryear{{Zheng} et~al.,}{{Zheng}
			et~al.}{2005}]{Zheng05}
		{Zheng} Z.,  et~al., 2005, \mn@doi [\apj] {10.1086/466510}, \href
		{http://adsabs.harvard.edu/abs/2005ApJ...633..791Z} {633, 791}
		
		\bibitem[\protect\citeauthoryear{{Zheng}, {Coil}  \& {Zehavi}}{{Zheng}
			et~al.}{2007}]{Zheng07}
		{Zheng} Z.,  {Coil} A.~L.,   {Zehavi} I.,  2007, \mn@doi [\apj]
		{10.1086/521074}, \href {http://adsabs.harvard.edu/abs/2007ApJ...667..760Z}
		{667, 760}
		
		\bibitem[\protect\citeauthoryear{{Zheng}, {Zehavi}, {Eisenstein}, {Weinberg}
			\& {Jing}}{{Zheng} et~al.}{2009}]{Zheng09}
		{Zheng} Z.,  {Zehavi} I.,  {Eisenstein} D.~J.,  {Weinberg} D.~H.,   {Jing}
		Y.~P.,  2009, \mn@doi [\apj] {10.1088/0004-637X/707/1/554}, \href
		{http://adsabs.harvard.edu/abs/2009ApJ...707..554Z} {707, 554}
		
		\bibitem[\protect\citeauthoryear{{Zhu}, {Zheng}, {Lin}, {Jing}, {Kang}  \&
			{Gao}}{{Zhu} et~al.}{2006}]{Zhu06}
		{Zhu} G.,  {Zheng} Z.,  {Lin} W.~P.,  {Jing} Y.~P.,  {Kang} X.,   {Gao} L.,
		2006, \mn@doi [\apjl] {10.1086/501501}, \href
		{http://adsabs.harvard.edu/abs/2006ApJ...639L...5Z} {639, L5}
		
		\bibitem[\protect\citeauthoryear{{Zu} \& {Mandelbaum}}{{Zu} \&
			{Mandelbaum}}{2015}]{Zu15}
		{Zu} Y.,  {Mandelbaum} R.,  2015, \mn@doi [\mnras] {10.1093/mnras/stv2062},
		\href {http://adsabs.harvard.edu/abs/2015MNRAS.454.1161Z} {454, 1161}
		
		\bibitem[\protect\citeauthoryear{{Zu}, {Zheng}, {Zhu}  \& {Jing}}{{Zu}
			et~al.}{2008}]{Zu08}
		{Zu} Y.,  {Zheng} Z.,  {Zhu} G.,   {Jing} Y.~P.,  2008, \mn@doi [\apj]
		{10.1086/591071}, \href {http://adsabs.harvard.edu/abs/2008ApJ...686...41Z}
		{686, 41}
		
		\bibitem[\protect\citeauthoryear{{van den Bosch}, {More}, {Cacciato}, {Mo}  \&
			{Yang}}{{van den Bosch} et~al.}{2013}]{Bosch13}
		{van den Bosch} F.~C.,  {More} S.,  {Cacciato} M.,  {Mo} H.,   {Yang} X.,
		2013, \mn@doi [\mnras] {10.1093/mnras/sts006}, \href
		{http://adsabs.harvard.edu/abs/2013MNRAS.430..725V} {430, 725}
		
		\makeatother
	\end{thebibliography}

\bsp	% typesetting comment
\label{lastpage}
\end{document}